\documentclass[onefignum,onetabnum,a4paper]{siamart190516_custom}
\pdfoutput=1

\usepackage{lipsum}
\usepackage{amsfonts,bm}
\usepackage{graphicx}
\usepackage{epstopdf}
\usepackage{enumerate}
\usepackage{subcaption}
\usepackage{tikz}
\usepackage{pgfplots}
\usepackage{tudacolors}
\usepackage{siunitx}

\usepackage[hmargin=3.99cm, vmargin=4.3725cm]{geometry}

\usepackage[labeled]{multibib}
\newcites{SM}{Additional references}

\ifpdf
  \DeclareGraphicsExtensions{.eps,.pdf,.png,.jpg}
\else
  \DeclareGraphicsExtensions{.eps}
\fi

\def\tit{Rational kernel-based interpolation for complex-valued frequency response functions}
\def\kw{Complex-valued kernel methods, dynamical systems, frequency response function, model selection, rational approximation}

\hypersetup{
  bookmarksnumbered=true,
  pdfauthor={Julien Bect, Niklas Georg, Ulrich Römer, Sebastian Schöps},
  pdftitle={\tit},
  pdfsubject={},
  pdfkeywords={\kw}}

\newsiamremark{remark}{Remark}
\newsiamremark{hypothesis}{Hypothesis}
\crefname{hypothesis}{Hypothesis}{Hypotheses}
\newsiamthm{claim}{Claim}

\headers{Kernel-based interpolation for complex-valued functions}{J. Bect, N. Georg, U. Römer and S. Schöps}

\title{\tit\thanks{%
  Accepted for publication in \href{https://epubs.siam.org/journal/sjoce3}{SIAM Journal of Scientific Computing}, %
  DOI:\href{https://dx.doi.org/10.1137/23M1588901}{10.1137/23M1588901}. %
  Author-generated version. %
  Preliminary results have been published in the PhD thesis of the second author \cite{phdGeorg}.
\funding{The work of N. Georg and U. Römer was funded by the Deutsche Forschungsgemeinschaft (DFG, German Research Foundation) -- RO4937/1-1. The work of N. Georg is also supported by the Graduate School CE within the Centre for Computational Engineering at Technische Universität Darmstadt.}}}

\author{
  Julien Bect\footnotemark[3]
\and Niklas Georg\footnotemark[4]
\and Ulrich Römer\footnotemark[2]
\and Sebastian Schöps\footnotemark[4]}

\usepackage{amsopn}

\DeclareMathOperator{\vect}{span}

\newcommand \Cset {\mathbb C}
\newcommand \Dset {\mathbb D}  
\newcommand \Kset {\mathbb K}  
\newcommand \Nset {\mathbb N}  
\newcommand \Rset {\mathbb R}
\newcommand \Xset {\mathbb X}
\newcommand \Sset {\mathbb S}

\newcommand \Esp  {\mathsf E}

\newcommand   \Acal {\mathcal{A}}
\renewcommand \i    {\mathrm{i}}    
\newcommand   \Hol  {\mathrm{Hol}}  

\renewcommand \subset {\subseteq}

\newcommand \re  {\mathrm{R}}    
\newcommand \im  {\mathrm{I}}    
\newcommand \LOO {\mathrm{LOO}}  

\newcommand{\tra}{{\mskip -1mu\scriptstyle\mathrm{T}}}
\newcommand{\her}{{\mskip -1mu\scriptstyle\mathrm{H}}}

\newcommand{\dt}{\mathrm{d}t}

\usepackage{circuitikz}
\usetikzlibrary{patterns}
\usetikzlibrary{arrows.meta}
\usetikzlibrary{shapes}
\usepgfplotslibrary{fillbetween}
\usepackage{tikz-3dplot}
\usetikzlibrary{plotmarks, positioning, backgrounds}

\begin{document}

\maketitle

\renewcommand{\thefootnote}{\fnsymbol{footnote}}
\footnotetext[2]{Institut für Akustik und Dynamik, %
  Technische Universität Braunschweig, Braunschweig, Germany %
  (\email{u.roemer@tu-braunschweig.de}).}
\footnotetext[3]{Université Paris-Saclay, CNRS, CentraleSupélec, %
  Laboratoire des signaux et systèmes, Gif-sur-Yvette, France %
  (\email{julien.bect@centralesupelec.fr}).}
\footnotetext[4]{Computational Electromagnetics Group, %
  Technische Universität Darmstadt, Darmstadt, Germany %
  (\email{georg@gsc.tu-darmstadt.de}, \email{sebastian.schoeps@tu-darmstadt.de}).}
\renewcommand{\thefootnote}{\arabic{footnote}}

\begin{abstract}
This work is concerned with the kernel-based approximation of a
complex-valued function from data, where the response
function of a partial differential equation in the frequency domain is
of particular interest. %
In this setting, kernel methods are employed more and more frequently,
however, standard kernels do not perform well. %
Moreover, the role and mathematical implications of the underlying
pair of kernels, which arises naturally in the complex-valued case, remain
to be addressed. %
We introduce new reproducing kernel Hilbert spaces of complex-valued functions, and formulate the
problem of complex-valued interpolation with a kernel pair as minimum
norm interpolation in these spaces. %
Moreover, we combine the interpolant with a low-order rational
function, where the order is adaptively selected based on a new model
selection criterion. %
Numerical results on examples from different fields, including electromagnetics
and acoustics examples, illustrate the performance of the method
in comparison to available rational approximation methods.


\end{abstract}



\section{Introduction}
We consider dynamical systems of the form
\begin{equation}
  \label{eq:generalDiscretePDEDynamic}
  \mathbf{M} \ddot{\mathbf{u}}(t) + \mathbf{D} \dot{\mathbf{u}}(t)
  + \mathbf{K}  \mathbf{u}(t) = \mathbf g(t),
\end{equation}
to be endowed with initial conditions and
$\mathbf K, \mathbf D, \mathbf M \in \mathbb R^{n_h\times n_h}$,
$\mathbf{u}(t),\mathbf{g}(t) \in \mathbb R^{n_h}$. %
We are in particular interested in approximating scalar time-dependent
quantities derived from the solution, of the form
\begin{equation}
  \label{eq:generalScalarQuantity}
  f(t) =  \mathbf{j}^\tra \mathbf{u}(t),
  \qquad \mathbf{j} \in \mathbb{R}^{n_h},
\end{equation}
which are commonly used to assess engineering designs. %
System \eqref{eq:generalDiscretePDEDynamic} may stem from a partial
differential equation after spatial discretization with $n_h$ degrees
of freedom. %
In a mechanics context, $\mathbf{K},\mathbf{D},\mathbf{M}$ are
referred to as stiffness, damping and mass matrix, but problems
arising in many areas of science and engineering can be brought into
this form. Our numerical results will cover electromagnetic and
acoustic field problems in particular.  In view of the linearity of
the equation, a frequency domain analysis is often adopted. %
Assuming for simplicity that~$\mathbf{u}$ and~$\dot{\mathbf{u}}$
vanish at~$t = 0$, the (one-sided) Laplace transform
of~\eqref{eq:generalDiscretePDEDynamic}--\eqref{eq:generalScalarQuantity}
with respect to the time variable~$t$ is
\begin{equation}
  \label{eq:generalDiscretePDE}
  \begin{split}
    \left( s^2 \mathbf{M}  + s \mathbf{D} + \mathbf{K} \right)\, \hat{\mathbf{u}}(s)
    &= \hat{\mathbf{g}}(s), \\
    \hat{f}(s) &= \mathbf{j}^\tra \mathbf{\hat{u}}(s),
  \end{split}
\end{equation}
where $s$ denotes the complex frequency variable, also known as the
Laplace variable. %
Assuming a suitably normalized excitation $\hat{\mathbf{g}}(s)$, the
frequency response function is defined as the value
$\omega \mapsto \hat f (\i \omega)$ of~$\hat f$ on the imaginary axis,
where $\omega$ is called the angular frequency, and we are typically
interested more specifically in its value on a certain interval
$\Omega = \left[ \omega_{\min},\omega_{\max} \right] \subseteq
[0,+\infty)$. %
In the following, we omit explicitly indicating frequency domain
variables to simplify the notation.

The location of the poles of $\hat f$ strongly depends on the
properties of $\mathbf{K},\mathbf{D},\mathbf{M}$, see
\cite{tisseur2001quadratic}. %
We assume, in particular, that no pole is placed on the frequency axis
$\i \mathbb{R}$ and that the frequency response function is
holomorphic on the shifted right
half-plane~$\Gamma_\alpha = \{s \in\mathbb C \,\mid
\,\Re[s]>-\alpha\},\,\alpha>0$. The real parts of all poles are strictly negative
for instance if $\mathbf{K},\mathbf{D},\mathbf{M}$ are symmetric
positive definite, see Section 3 of \cite{tisseur2001quadratic}. %
The same holds true if the homogeneous version of
\eqref{eq:generalDiscretePDEDynamic} is stable, in the sense that all
solutions decay exponentially to zero as $t \rightarrow \infty$. %
The holomorphy of response functions has recently been studied also in
the context of partial differential equations, see \cite[Proposition
5.3]{Bonizzoni_2020aa} for instance. %
There, the frequency response map for an acoustic scattering problem
was studied and appropriate damping terms ensured a locally
holomorphic response function, with a negative real part for all
poles\footnote{Because of a different convention
  \cite{Bonizzoni_2020aa} establishes a negative \emph{imaginary} part
  of the eigenvalues}.%

\begin{remark}
  The method introduced in this paper was motivated by problems of the
  form~\eqref{eq:generalDiscretePDEDynamic}, but can be readily
  applied to the approximation of the frequency response function of
  any linear, time-invariant and asymptotically stable dynamical
  system.
\end{remark}

Adopting a data-driven approach, ~\eqref{eq:generalDiscretePDE} must
be solved repeatedly on a set of interpolation/training points
$\omega_i \in \Omega$, with $s_i = \i \omega_i$. %
Numerical efficiency demands a small training set
\begin{equation}
  \left( \omega_i,\, f(\i\omega_i) \right)_{1 \le i \le n},
  \quad
  \mathrm{where}~\omega_i\in\Omega,\, f(\i\omega_i)\in\mathbb C,\,
  i=1, \ldots, n.\label{eq:training_set}
\end{equation} 
Hence, there is a need for accurate interpolation in the frequency
domain.

The data-driven approximation of frequency response functions has
attracted considerable interest in the literature, see for instance
\cite{gustavsen1999rational,lataire2016transfer, nakatsukasa2018aaa}
and the references therein. %
Among the numerous available approaches we mention vector fitting
\cite{gustavsen1999rational} and the adaptive Antoulas-Anderson (AAA) method
\cite{nakatsukasa2018aaa} in particular, which are widely used,
state-of-the-art approximation methods.

Vector Fitting (VF) is a rational approximation technique,
specifically tailored to functions in the frequency domain. %
It is based on a representation in terms of partial fractions as
\begin{equation}
  f(\i\omega) \approx \sum_{m=1}^{M} \frac{r_m} {\i\omega-p_m}
  + d +\i\omega h,\label{eq:part_frac}
\end{equation}
where the $M$ poles $p_m$ are relocated in each iteration by solving a
linear least-square problem, see
\cite{gustavsen2006improving,gustavsen1999rational} for details. %
The implementation guarantees that all poles are stable,
i.e. $\mathcal R[p_m]<0$, and are either real or come in
complex-conjugate pairs. %

The AAA method \cite{nakatsukasa2018aaa}
employs the barycentric interpolation
\begin{equation}
  \label{eq:AAA}
  f(\i\omega) \approx  r(\omega)
  = \frac{n(\omega)}{d(\omega)}
  = \frac{\sum_{j\in J}{\frac {w_jf(\i \omega_j)}{\omega-\omega_j}}}{\sum_{j\in J} \frac{w_j}{\omega-\omega_j}},
\end{equation} 
where $J \subseteq \{1, \ldots, n\}$ has cardinality~$m$. %
The rational function in \eqref{eq:AAA} is of type $(m-1,m-1)$, which
can be seen by multiplying both numerator and denominator by
$\prod_{j\in J} (\omega-\omega_j)$. %
Moreover, $r(\omega_j) = f(\i \omega_j)$ for all $j\in J$. %
The weights~$w_j$ and nodes~$\omega_j$, $j\in J$, are determined
adaptively in a two-step procedure, based on linear least squares
problems and a greedy strategy \cite{nakatsukasa2018aaa}.

Other data-driven approaches, related to rational interpolation and
model order reduction are the Loewner framework
\cite{antoulas2017tutorial} and the recent contribution
\cite{nobile2020non}, which employs the Heaviside representation. 
A Bayesian rational Polynomial Chaos-type model has been put forth in \cite{schneider2023sparse} to capture the effect of uncertain parameters, e.g., on frequency response functions. 
A complex-valued version of support vector machine regression has been
presented in \cite{treviso2021multiple}, which is restricted to the
so-called circular case with a single kernel only. %
Complex interpolation with a pair of kernels has been addressed in
\cite{boloix2017widely,picinbono1995widely} and also from a Gaussian
process regression perspective in \cite{boloix2018complex,
  hallemans2022frf}. 

Despite recent progress with complex kernel methods, a general
framework with a complete mathematical background on the underlying
reproducing kernel Hilbert spaces (RKHS) is missing. %
In comparison to parametric rational approximation methods, e.g., AAA
and vector fitting, the kernel/Gaussian process approach is appealing
because of its principled statistical foundations, which allow for
model selection, uncertainty quantification and adaptive sampling. %
Additionally, desired properties of the system, such as stability and
causality, can be ensured during kernel design \cite{hallemans2022frf}. %
Adaptive sampling, in particular, is more involved for Loewner-type approaches. In the standard formulation of AAA, for instance, new support points are chosen from a discrete set of a priori fixed points. %
An exception is the recently introduced Greedy-type adaptive sampling
Loewner approach in \cite{pradovera2023toward}.%
 
In this paper, we introduce a new kernel-based interpolation method
which is well adapted to frequency responses. %
We will put special emphasis on the complex-valued setting and show
that the data are used more efficiently if a dedicated kernel method
is constructed and interpolation of the real and imaginary part
individually is avoided. %
To address problems with a few dominant poles we include a low-order
rational basis into the kernel method and present a new model
selection scheme. %
We compare our rational kernel-based interpolation method against both
AAA and vector fitting and observe an improved or at least comparable
performance for a variety of test cases. %
Finally, the paper develops the required notions of RKHS
and minimum norm interpolation for complex-valued kernel methods in
general.

The material is structured in the following way. %
In \Cref{sec:theory} we introduce the concept of a complex/real
kernel Hilbert space and consider the special case of frequency
response functions as well as the connections to complex-valued
Gaussian process regression. %
\Cref{sec:alg} introduces our new method, which employs a
kernel, a pseudo-kernel and an additional rational basis for capturing
dominant poles. %
Finally, \Cref{sec:numerics} reports several examples from
PDE-based applications, comparing our method to AAA and vector fitting
before conclusions are drawn.

\medbreak\itshape

Nota bene: %
A method sharing some similarities with the one proposed in
Section~\ref{sec:alg} has been published recently in the automatic
control literature \cite{hallemans2022frf}. %
We became aware of it at very late stage in the writing of the present
article. %
After introducing our new method in Section~\ref{sec:alg}, we discuss
similarities and differences in Remark~\ref{rmk:hallemans}.

\medbreak\normalshape

\section{Complex/Real RKHS interpolation}
\label{sec:theory}
In order to address kernel-based interpolation of the frequency
response function, we start by recalling basic facts on RKHSs; %
see, e.g., \cite{Paulsen_2016aa} for a comprehensive introduction to
this topic. %
Then, new results establishing the theoretical basis of our method
are stated, the proofs of which can be found in
Appendix~\ref{sec:proofs-appendix}.

\begin{definition}[Complex RKHS] \label{def:complex-rkhs}
  A complex RKHS~$H$ over a non-empty set~$\Sset$ is a complex Hilbert
  space of functions~$\Sset\rightarrow\mathbb C$ such that, for all
  $s \in \Sset$, the evaluation functional
  $\delta_s: H \rightarrow \Cset$, $f \mapsto f(s)$, is continuous.
\end{definition}

The Riesz representation theorem implies that there exists a unique
function $k:\Sset\times \Sset \rightarrow \Cset$, called the
reproducing kernel of~$H$, such that $k(\cdot, s) \in H$ and
\begin{equation} \label{equ:reproduction-property}
  f(s) = \delta_s(f) = \left< f, k(\cdot,s) \right>_H
\end{equation}
for all~$s\in \Sset$ and~$f \in H$, %
where $\left< \cdot,\cdot \right>_H$ denotes the Hermitian inner
product of~$H$. %
Equation~\eqref{equ:reproduction-property} is called the reproduction
property, and it is easily seen that the kernel~$k$ is
\emph{Hermitian} %
(i.e., $k(s, s_0) = k(s_0, s)^*$ for all~$s, s_0 \in \Sset$) and
\emph{positive definite}: for all $n \in \Nset^*$ and all~$(s_1, \alpha_1)$,
\ldots, $(s_n, \alpha_n) \in \Sset \times \Cset$,
\begin{equation}
	\label{eq:kernel_positiveDefinite}
  \sum_{1 \le i, j\le n} \alpha_i^* \alpha_j k(s_i, s_j) \;\ge\; 0.
\end{equation}

\begin{theorem}[Moore-Aronszajn]
	\label{thm:Moore-Aronszajn}
  For any positive definite Hermitian
  kernel~$k: \Sset\times \Sset \rightarrow \Cset$, there exists a
  unique complex Hilbert space~$H$ of functions on~$\Sset$ such that
  the reproduction property holds with reproducing kernel~$k$.
\end{theorem}

Real RKHSs are defined similarly, replacing~$\Cset$ by~$\Rset$ in
Definition~\ref{def:complex-rkhs}: %
in this case $H$~is a real Hilbert space, the reproducing kernel is
symmetric positive definite, and a suitably modified statement of the
Moore-Aronszajn theorem holds as well.

\begin{theorem}[Interpolation] %
  \label{thm:complexRKHSinterpolation} %
  Let $H$ be a real or complex RKHS over~$\Sset$ %
  with kernel $k: \Sset \times \Sset \to \Kset$, %
  where $\Kset = \Rset$ or~$\Cset$ depending on the type of RKHS. %
  Let $n \in \Nset^*$, $s_1, \ldots, s_n \in \Sset$ %
  and $y_1, \ldots, y_n \in \Kset$. %
  Then there exists a function $g \in H$ such that $g(s_i)=y_i$ for
  all $i \in \{ 1, \ldots, n \}$ if, and only if, the system
  \begin{equation}\label{equ:interp-system}
    \begin{bmatrix}
      k(s_1,s_1) &\dots & k(s_1,s_n)\\
      \vdots & \ddots & \vdots\\
      k(s_n,s_1) &\dots & k(s_n,s_n)
    \end{bmatrix}
    \begin{bmatrix}
      \gamma_1\\
      \vdots\\
      \gamma_{n}
    \end{bmatrix}
    = \begin{bmatrix}
      y_1\\\vdots\\y_n
    \end{bmatrix}
  \end{equation}
  admits a solution. %
  Furthermore, for any solution of~\eqref{equ:interp-system},
  $g = \sum_{i=1}^n \gamma_i\, k(\cdot,s_i)$ is the unique interpolant
  of the data $\left( s_1, y_1 \right)$, \ldots,
  $\left( s_n, y_n \right)$ with minimal norm in~$H$.
\end{theorem}
A positive definite kernel is called \emph{strictly} positive definite
if the kernel matrix $K_n=(k(s_i,s_j))_{1\leq i,j\leq n}$ is
invertible (equivalently, if \eqref{eq:kernel_positiveDefinite}~is
strict for all~$\left( \alpha_1, \ldots, \alpha_n \right) \neq 0$)
whenever $s_1,\ldots,s_n$ are distinct points. %
This ensures that \eqref{equ:interp-system} has a unique solution.

We will proceed by introducing several complex RKHS and their
kernels. For $s \in \mathbb C$, let $\Re[s]$ and $\Im[s]$ denote the
real and imaginary part, respectively. %
An important example is the Hardy space $H^2(D)$ on the unit disc $D=\{s \in \Cset: |s| <1 \}$.
This space plays a role in the analysis of the stability of discrete
dynamical systems, see \cite{baratchart1991identification}, for
instance.  
Here, in the context of continuous-time dynamical systems, we are more
interested in the corresponding Hardy space 
\begin{equation}
  H^2(\Gamma_\alpha) \!=\!
  \left\{
    f\in \Hol(\Gamma_\alpha)\!:\!\|f\|_{H^2(\Gamma_\alpha)}
    = \sup_{x>-\alpha}\left(\int_{-\infty}^{\infty} \left| f(x + \i y)^2 \right|\,\mathrm d y \right)^{\frac 1 2} <\infty
  \right\},
\end{equation}
where $\Hol(\Gamma_\alpha)$ denotes the space of holomorphic functions
on~$\Gamma_\alpha$. %
Note, that there is a Banach space isometry between the $H^2$ spaces on disc and half-plane, see \cite[Chapter
8]{hoffman:1962} for details. %
\begin{theorem}\label{thm:gammaAlpha}
  The space $H^2(\Gamma_\alpha)$ is a complex RKHS, with strictly
  positive definite reproducing kernel~$k$ given by
  \begin{equation}\label{equ:szego} 
    k_\alpha\left( s, s_0 \right)
    = \frac{1}{2\pi\, \left( 2\alpha + s + s_0^*\right)},
    \quad s, s_0\in\Gamma_{\alpha}.
  \end{equation}
\end{theorem}

Following standard terminology in complex analysis (see, e.g., \cite{krantz:2001:function}), we will refer
to~$k_{\alpha}$ as the \emph{Szegö kernel} for the
domain~$\Gamma_\alpha$. %
Evaluating \eqref{equ:szego} only on the imaginary axis $s=\i \omega$,
the expression simplifies to
\begin{equation}
  k_\alpha\left( \i\omega, \i\omega_0 \right) =
  \frac 1 {2\pi\left( 2\alpha+\i(\omega - \omega_0) \right)},
  \quad \omega, \omega_0 \in \Omega.\label{eq:SzegoFrequency}
\end{equation}

We consider the stable spline kernel
\cite{pillonetto2010new,lataire2016transfer} as another example. This
kernel has been proposed in the time domain to model functions with a
certain smoothness, which additionally incorporate impulse response
stability \cite{pillonetto2010new}. The corresponding kernel for the
frequency domain transfer function has been obtained in
\cite{lataire2016transfer} and reads
\begin{multline}
  \label{eq:stablespline}
  k_\alpha\left( \i\omega, \i\omega_0 \right) =
  \frac{1}{2} \frac{1}{3 \alpha + \i (\omega - \omega_0)} \times \\
  \left( %
    \frac{1}{2 \alpha + \i \omega} + \frac{1}{2 \alpha - \i \omega_0} %
    - \frac{1}{3(3 \alpha + \i \omega)} - \frac{1}{3(3 \alpha - \i \omega_0)} %
  \right).
\end{multline}
Other related kernels can be found in the control literature, see \cite{lataire2016transfer,hallemans2022frf}.

\subsection{Complex/real RKHS interpolation}
\label{sec:complex-real-RKHS}

The frequency response function fulfills the symmetry property
$f^*(s) = f(s^*)$ for all~$s \in \Gamma_\alpha$, since it is the
Laplace transform of a real-valued function. %
We are thus naturally led to cast our interpolation problem not
in~$H^2(\Gamma_{\alpha})$ but in the subset
\begin{equation} \label{equ:H2sym}
  H^2_\mathrm{sym}(\Gamma_{\alpha}) \;=\; \bigl\{
  f\in H^2(\Gamma_{\alpha}):\; \forall s\in\Gamma_\alpha,\; f^*(s)=f(s^*)
  \bigr\}.
\end{equation}
This set of complex-valued functions, however, cannot by endowed with
the structure of a complex RKHS. %
In fact, it is not even a vector space over~$\Cset$: %
indeed, for any $f\in H^2_\mathrm{sym}(\Gamma_\alpha)$ and
$s\in \Gamma_{\alpha}$, we would have
$(\i f)^*(s) = -\i f^*(s) = -\i f(s^*)$ and
$(\i f)^*(s) = (\i f)(s^*) = \i f(s^*)$, %
which is a contradiction if $f(s^*) \neq 0$.

Observing that the subset of~$H^2(\Gamma_{\alpha})$ defined
by~\eqref{equ:H2sym} is a real vector space of complex-valued functions,
we define in the following a new type of function space, which we call
a complex/real RKHS.
\begin{definition}[Complex/real RKHS] \label{def:CRrkhs}%
  Let $\Sset$ denote a non-empty set and let $H$ denote a real Hilbert
  space of complex-valued functions on~$\Sset$. %
  We say that $H$ is a complex/real RKHS if the evaluation functionals
  are continuous (i.e., for all $s \in \Sset$, the function
  $\delta_s: H \to \Cset$, $f \mapsto f(s)$, is continuous).
\end{definition}

In the remaining part of this section we will establish general
results related to these
spaces. Section~\ref{sec:complex-real-RKHS-sym} will then present
consequences for the RKHS with the symmetry property
$f^*(s) = f(s^*)$.

\begin{remark} \label{rem:complex-RKHS-subspaces}
  Any complex RKHS~$H$ (such as $H^2(\Gamma_{\alpha})$) can be seen as a
  complex/real RKHS by forgetting the complex structure, i.e., by
  considering $H$ as a real vector space, endowed with the real inner
  product $\left<f, g\right> \mapsto \Re \left( \left< f, g \right>_H \right)$. %
  More generally, any real subspace of~$H$ (such
  as~$H^2_\mathrm{sym}(\Gamma_{\alpha})$), endowed with this inner
  product, is clearly a complex/real RKHS. %
  The converse statement is false, however.
\end{remark}

\begin{proposition} \label{prop:counterexample-dim2}
  There exists a complex/real RKHS of dimension two over the reals
  that is not a real subspace of a complex RKHS.
\end{proposition}

The elements of a complex/real RKHS are complex-valued functions
over~$\Sset$, but can be conveniently represented as real-valued
functions over $\tilde\Sset = \Sset \times \{\re,\im\}$ through the
mapping $\Acal: \Cset^\Sset \rightarrow \Rset^{\tilde\Sset}$ defined
by
\begin{equation} \label{eq:mappingNonIntrusive}
  (\Acal f)(s,a) = G_a(f(s)), 
\end{equation}
where $G_\re(s) = \Re(s)$ and $G_\im(s) = \Im(s)$. %
This mapping defines an isometric isomorphism of real Hilbert spaces
between $H$ and the real vector space
$\tilde H = \Acal H \subset \Rset^{\tilde\Sset}$, endowed with the
image inner product. %
The image space $\tilde H$ is easily seen to be a real RKHS if and
only if~$H$ is a complex/real RKHS: %
this observation will be useful both from a theoretical point of view,
to establish properties of complex/real RKHSs, and from a practical
point of view
(see Section~\ref{sec:numerics}).

\begin{remark}
  Complex/real RKHSs can also been seen a special case of
  vector-valued RKHSs~\cite{burbea:1984:banach,
    micchelli:2005:vector}, through the usual identification
  of~$\Cset$ with~$\Rset^2$.
\end{remark}

The term ``functional'' is used in a loose sense in
Definition~\ref{def:CRrkhs}, since $H$ is a real vector space
while~$\delta_s$ is a complex-valued function. %
Therefore, in contrast with the usual case of complex RKHSs, the
continuous functionals~$\delta_s$, $s \in \Sset$, do not belong to the
topological dual of~$H$. %
The real and imaginary evaluation functions however---namely,
$\Re \circ \delta_s$ and~$\Im \circ \delta_s$---do belong to the
topological dual, and can thus be expressed through inner products.

\begin{proposition} \label{prop:repr-eval-func} %
  Let $H$ be a complex/real RKHS on a set~$\Sset$, and set %
  \begin{equation}
    k_{a\mskip 1mu a_0}(s, s_0) \;=\; \tilde k\left( (s, a),\, (s_0, a_0) \right),
    \qquad s, s_0 \in \Sset,
    \quad a, a_0 \in \{ \re, \im \},
  \end{equation}
  where $\tilde k$ denotes the reproducing kernel of~$\tilde H = \Acal H$. %
  Then, for all $s \in \Sset$, we have
  \begin{equation}
    \delta_s \;=\;
    \underbrace{\left<\, \bm{\cdot}\,,\, \varphi_\re(\cdot,s) \right>_H}_{\Re \circ\, \delta_s}
    \;+\; \i \underbrace{\left<\, \bm{\cdot}\,,\, \varphi_\im(\cdot,s) \right>_H}_{\Im \circ\, \delta_s},
  \end{equation}
  where %
  $\varphi_\re = k_{\re\re} \,+\, \i\, k_{\im\re}$ and %
  $\varphi_\im = k_{\re\im} \,+\, \i\, k_{\im\im}$.
\end{proposition}

This result associates to each complex/real RKHS a
pair~$\left( \varphi_\re, \varphi_\im \right)$ of kernels
$\varphi_{a}:\Sset \times \Sset \to \Cset$, $a \in \{ \re, \im\}$. %
Characterizing admissible choices for this pair of kernels, in the
spirit of Theorem~\ref{thm:Moore-Aronszajn} for complex RKHSs, %
is possible but not convenient. %
Instead, motivated by the connection between complex/real RKHSs and
complex Gaussian processes (to be discussed in
Section~\ref{sec:relation-GPs}), and in particular the work of
Picinbono \cite{picinbono:1996}, we introduce another pair of kernels
as follows.

\begin{definition} \label{def:complex-kernels} %
  Let $H$ denote a complex/real RKHS and let $k_{\re\re}$, $k_{\im\im}$,
  $k_{\re\im}$, $k_{\im\re}$, $\varphi_\re$ and~$\varphi_\im$ be defined as in
  Proposition~\ref{prop:repr-eval-func}. %
  Then we define the \emph{complex kernel}~$k$ of the complex/real
  RKHS as
  \begin{equation}
    \label{eq:complex-kernel}
    k %
    \;=\;  \left( k_{\re\re} + k_{\im\im} \right) %
    \,+\, \i\, \left( k_{\im\re} - k_{\re\im} \right)
    \;=\; \varphi_\re - \i \varphi_\im,
  \end{equation}
  and its \emph{pseudo-kernel} $c$ as:
  \begin{equation}
    \label{eq:pseudo-kernel}
    c %
    \;=\; \left( k_{\re\re} - k_{\im\im} \right) %
    \,+\, \i\, \left( k_{\im\re} + k_{\re\im} \right)
    \;=\; \varphi_\re + \i \varphi_\im.
  \end{equation}
\end{definition}

\begin{proposition}\label{prop:dense-subspace}
  The functions of the form
  $\gamma\, k(\cdot,s_0) + \gamma^*\, c(\cdot, s_0)$, with
  $\gamma \in \Cset$ and $s_0 \in \Sset$, span a dense subset of~$H$.
\end{proposition}

\begin{remark}
  Proposition~\ref{prop:dense-subspace} suggests that the concept of a
  complex/real RKHS, introduced in this article, provides a rigorous
  formalization of the idea of a ``wide-linear complex-valued RKHS''
  (WL-RKHS) proposed in~\cite{boloix2017widely} (see Definition~3.1).
\end{remark}

It can be shown that the complex/real RKHS obtained by forgetting the
complex structure of a complex RKHS with reproducing kernel~$k_0$, as
described in Remark~\ref{rem:complex-RKHS-subspaces}, is the
complex/real RKHS with complex kernel $k = 2 k_0$ and vanishing
pseudo-kernel---which, borrowing terminology from the signal
processing literature \cite{picinbono:1996}, can be called
\emph{circular}. %
The factor~$2$ in the relation between~$k$ and~$k_0$ is the price to
pay for the consistency of Definition~\ref{def:complex-kernels} with
the concepts of covariance and pseudo-covariance functions for complex
Gaussian processes (see Section~\ref{sec:relation-GPs}). %
More generally, we have the following characterization of the set of
admissible $(k, c)$ pairs.

\begin{theorem} \label{thm:CR-RKHS-characterization}\setlength{\parskip}{2pt}%
  For a given complex/real RKHS~$H$, the kernels~$k$ and~$c$
  introduced in Definition~\ref{def:complex-kernels} satisfy the
  following:
  \begin{enumerate}[i) ]
  \item $k$ is complex-valued, Hermitian and positive definite.
  \item $c$ is complex-valued and symmetric.
  \end{enumerate}
  Moreover, for all~$n \ge 1$ and all~$s_1, \ldots s_n \in \Sset$:
  \begin{enumerate}[i) ] \setcounter{enumi}{2}
  \item $\ker K_n \subset \ker C_n^*$ and,
  \item if $K_n$ is positive definite,
    $K_n^* - C_n^* K_n^{-1} C_n$ is positive semi-definite,
  \end{enumerate}
  where $K_n = \left( k(s_i,s_j) \right)_{1 \le i,j \le n}$ and
  $C_n = \left( c(s_i,s_j) \right)_{1 \le i,j \le n}$.

  Conversely, for any pair of functions
  $k, c:\Sset \times \Sset \to \Cset$ that satisfies these four
  properties, there exists a unique complex/real RKHS on~$\Sset$ with
  complex kernel~$k$ and pseudo-kernel~$c$.
\end{theorem}

\begin{theorem}[Interpolation in a complex/real RKHS] %
  \label{thm:interp:cr} %
  Let $H$ denote a complex/real RKHS over~$\Sset$ %
  with complex kernel~$k$ and pseudo-kernel~$c$. %
  Let $n \in \Nset^*$, $s_1, \ldots, s_n \in \Sset$ %
  and $y_1, \ldots, y_n \in \Cset$. %
  Then there exists a function $g \in H$ such that $g(s_i)=y_i$ for
  all $i \in \{ 1, \ldots, n \}$ if, and only if, the system
  \begin{equation}
    \label{equ:interp-system:cr}
    K_n \gamma + C_n \gamma^* = y
  \end{equation}
  admits a solution $\gamma \in \Cset^n$, where
  $K_n = \left( k(s_i,s_j) \right)_{1 \le i,j \le n}$,
  $C_n = \left( c(s_i,s_j) \right)_{1 \le i,j \le n}$,
  and $y = \left( y_1, \ldots, y_n \right)^\tra$. %
  Furthermore, for any solution of~\eqref{equ:interp-system:cr},
  \begin{equation}
    \label{equ:interpolant:cr}
    g = \sum_{i=1}^n \gamma_i\, k(\cdot,s_i)
    + \sum_{i=1}^n \gamma_i^*\, c(\cdot,s_i)
  \end{equation}
  is the unique interpolant of the data $\left( s_1, y_1 \right)$,
  \ldots, $\left( s_n, y_n \right)$ with minimal norm in~$H$.
\end{theorem}

For the usual setting of real or complex RKHSs, strictly positive
definite kernels guarantee that the interpolation
system~\eqref{equ:interp-system} has a solution for any
data~$y_1, \ldots, y_n$. %
This remains true for the system~\eqref{equ:interp-system:cr} in the
case of a complex/real RKHS if the associated real kernel~$\tilde k$
is strictly positive definite on
$\tilde\Sset = \Sset \times \{\re,\im\}$.


\subsection{Complex/real RKHS with symmetry condition} \label{sec:complex-real-RKHS-sym}

We now characterize, in full generality, the complex/real
RKHSs where a symmetry condition of the form $f^*(s) = f(s^*)$ holds
for all~$f \in H$ and~$s \in \Sset$. %
The following theorem provides a necessary and sufficient condition on~$k$
for such a space to exist and gives the expression of the
corresponding pseudo-kernel. %
The expression appeared previously in \cite[Equations~(48)--(49)]{lataire2016transfer}
for a special type of kernel.

\begin{theorem}\label{thm:hermitian}
  Let $\Sset$ denote a non-empty set, equipped with an involution
  $s \mapsto s^*$ and $k:\Sset \times \Sset \to \Cset$ denote a
  Hermitian positive definite kernel on~$\Sset$. %
  Then the following assertions are equivalent:
  \begin{enumerate}[i) ]
  \item There exists a complex/real RKHS $H$ on~$\Sset$, with complex
    kernel~$k$, such that
    \begin{equation}
      \label{equ:symmetry}
      \forall f \in H,\; \forall s \in \Sset,\quad f^*(s) = f(s^*).
    \end{equation}
  \item There exists a complex/real RKHS $H$ on~$\Sset$, with complex
    kernel~$k$ and pseudo-kernel $c$ defined by
    \begin{equation}
      \label{equ:pseudo-kern-symm}
      \forall s,s_0 \in \Sset,\quad %
      c(s,s_0) = k(s, s_0^*).
    \end{equation}
  \item $\forall s,s_0 \in \Sset$, $k(s, s_0^*) = k(s_0, s^*)$.
  \end{enumerate}

  \smallbreak
  
  If any (and consequently all) of these assertions holds, then the
  complex/real RKHS~$H$ with complex covariance~$k$ and pseudo
  kernel~\eqref{equ:pseudo-kern-symm} is the unique RKHS on~$\Sset$
  with complex covariance~$k$ such that~\eqref{equ:symmetry} holds. %
  Moreover, denoting by $H_\Cset$ the complex RKHS with kernel~$k$, we
  have $H_\Cset = H \oplus \i H$,
  $H = \left\{ f \in H_\Cset \mid \text{\eqref{equ:symmetry} holds}
  \right\}$ and $\left< f, g \right> = \Re \left< f, g \right>_{H_\Cset}$
  for all $f, g \in H$.
\end{theorem}

It follows from this theorem that $H^2_\mathrm{sym}(\Gamma_{\alpha})$
can be characterized as the complex/real RKHS over~$\Gamma_\alpha$
with complex kernel~\eqref{equ:szego} and pseudo-kernel:
\begin{equation}\label{equ:pseudo-szego} 
  c_\alpha\left( s, s_0 \right) = \frac 1 {2 \pi (2\alpha + s + s_0)},\quad s,\, s_0\in\Gamma_{\alpha}.
\end{equation}

More generally, Theorem~\ref{thm:hermitian} shows that the problem of
minimum-norm interpolation in a complex RKHS, with a symmetry
constraint of the form~\eqref{equ:symmetry}, can be solved by
considering the equivalent problem of minimal-norm interpolation in
the complex/real RKHS with the same complex kernel and the
pseudo-kernel given by~\eqref{equ:pseudo-kern-symm}.
In presence of the symmetry condition, even if the complex kernel $k$
is strictly positive definite, $\tilde{k}$ is not and an additional
condition on the data is required to ensure that~\eqref{equ:interp-system}
has a solution.

\begin{theorem}\label{thm:existence-uniqueness-hermit}%
  In the setting of Theorem~\ref{thm:hermitian}, assume that $k$ is
  strictly positive definite, $c$~is given
  by~\eqref{equ:pseudo-kern-symm}, and $s_1, \ldots, s_n \in \Sset$
  are distinct. %
  Then \eqref{equ:interp-system:cr}~has a solution if, and only if,
  $y_j = y_i^*$ for all $i, j$ such that $s_j = s_i^*$. %
  When this holds, there is a unique solution such that
  $\gamma_i = \gamma_j^*$ for all $i, j$ such that~$s_j = s_i^*$.
\end{theorem}

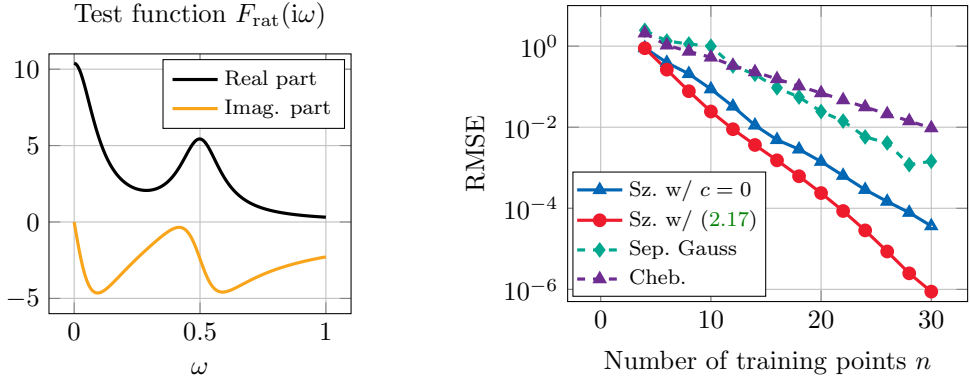
\begin{figure}
  \begin{subfigure}[t]{.43\textwidth}

    \begin{tikzpicture}
      \begin{axis}[grid, xlabel={$\omega$}, title={Test function
          $F_\mathrm{rat}(\i \omega)$}, width=1\textwidth,
        height=.9\textwidth, ymin=-6, ymax=11, ylabel near ticks,
        legend style = {font=\footnotesize}, legend cell align={left}]
        \addplot[black, very thick, no marks] table[x=x, y=y_ref_real, col sep = comma]{images/illustration_LowOrderRat1.csv};
        \addplot[TUDa-7b, very thick, no marks] table[x=x, y=y_ref_imag, col sep = comma]{images/illustration_LowOrderRat1.csv};
        \legend{Real part, Imag. part}
      \end{axis}
    \end{tikzpicture}

    \caption{Function $F_\mathrm{rat}(\i \omega)$ %
      (with real-valued inverse Laplace transform).}

  \end{subfigure}\hfill
  \begin{subfigure}[t]{.53\textwidth}

    \begin{tikzpicture}
      \begin{semilogyaxis}[grid, xlabel={Number of training points
          $n$}, ylabel={RMSE}, width=1\textwidth, height=.8\textwidth,
        legend pos = south west,xmin=-3,ymin=5e-7, ymax=10, %
        legend style = {font=\footnotesize, inner sep=1pt, outer sep=2pt, at=({0,0}), legend cell align={left}}]
        \addplot[TUDa-1b, very thick, mark=triangle, mark options={solid, fill=TUDa-1b}] table[x=N, y=RMSE, col sep = comma]{images/LowOrderRat1_Szego0P.csv};
        \addplot[TUDa-9b, very thick, mark=*, mark options={solid, fill=TUDa-9b}] table[x=N, y=RMSE, col sep = comma]{images/LowOrderRat1_Szego.csv};
        \addplot[TUDa-3b, very thick, dashed, mark=diamond*, mark options={solid, fill=TUDa-3b}] table[x=N, y=RMSE, col sep = comma]{images/LowOrderRat1_GaussSep.csv};
        \addplot[TUDa-11b, very thick, dashed, mark=triangle*, mark options={solid, fill=TUDa-11b}] table[x=N, y=RMSE, col sep = comma]{images/LowOrderRat1_Chebyshev.csv};
        \legend{%
          {Sz.\ w/~$c=0$}, %
          {Sz.\ w/~\eqref{equ:pseudo-kern-symm}}, %
          Sep.\ Gauss, Cheb.}
      \end{semilogyaxis}
    \end{tikzpicture}

    \caption{Convergence study of the RMSE for different
      approximations of $F_{\mathrm{rat}}(\i \omega)$.}

  \end{subfigure}

  \caption{%
    Left: Illustrations with the test function
    $F_{\mathrm{rat}} \in H^2_\mathrm{sym}(\Gamma_{0.1 + \epsilon}), \epsilon >0$ defined in
    \eqref{eq:Frat}. %
    Right: Convergence of the RMSE as a function of the number of
    (equidistant) training points. %
    Solid lines: complex/real interpolation with the Szegö kernel
    for~$H^2(\Gamma_\alpha)$, combined with the zero pseudo-kernel
    (blue) and the pseudo-kernel~\eqref{equ:pseudo-kern-symm} (red). %
    Dashed lines: interpolation with a Gaussian kernel for the real
    and imaginary part separately (green) and polynomial interpolation
    on Chebyshev nodes (purple).%
  }%
  \label{fig:PCov_demo}
\end{figure}

For illustration, we consider the third order rational function
\begin{equation}\label{eq:Frat}
  F_{\mathrm{rat}}(\i\omega)
  = \frac 1 {\i\omega-(-0.1)}
  + \frac {0.5}{\i\omega-(-0.1-0.5\i)}
  + \frac {0.5}{\i\omega-(-0.1+0.5\i)},\;
  \omega\in[0,1],
\end{equation}
which is the Laplace transform of the real-valued
function $t \mapsto e^{-0.1t}\bigl(1+\cos(0.5t)\bigr)$
and thus belongs to
$H^2_\mathrm{sym}(\Gamma_{0.1+\epsilon}) \subset H^2(\Gamma_{0.1+\epsilon})$ for all~$\epsilon > 0$. %
To illustrate the importance of the choice of pseudo-kernel,
we conduct a convergence study in
terms of the root-mean-square error (RMSE) of the approximations,
using equidistant training points
(details on the implementation and selection of
hyper-parameters will be given in the following sections). %
In Figure~\ref{fig:PCov_demo} we demonstrate that choosing a suitable
pseudo-kernel might have a significant impact on the convergence
properties of the (complex/real) RKHS interpolation. %
For the test function~\eqref{eq:Frat}, the pseudo-kernel~\eqref{equ:pseudo-kern-symm}
improves the convergence significantly. %
Note that the test function is a low order rational function which is
here only used to illustrate the impact of the pseudo-kernel. %
Accordingly, rational interpolation techniques as AAA or VF reach
machine accuracy already with $\approx8$ training points and are hence
excluded in the convergence plot for clarity. %
However, it can already be observed that complex/real RKHS
interpolation with the Szegö kernel outperforms the alternative
approach of separate kernel approximations for real and imaginary
part with a Gaussian kernel, as well as polynomial interpolation on
Chebyshev nodes.

\subsection{Relation to Gaussian process interpolation}
\label{sec:relation-GPs}
This section draws connections between minimum norm interpolation in a
RKHS and the posterior mean prediction of a Gaussian process (GP), for
both the complex and complex/real case. GPs are widely used, but to
the authors knowledge this is the first time that the RKHS associated
to any complex GP prediction is characterized. %
Another intention of this section is to make results from the GP
literature available for interpolation with a complex/real RKHS. %
In particular, we are interested in employing statistical methods for
model selection (see, e.g., \cite{petit:2023:parameter} and references
therein)---this will be further developed in
Section~\ref{sec:adap_proc}. %
We consider zero-mean processes in this section, %
for simplicity 
see Remark~\ref{rem:semi-RKHS} below.

\newcommand \kGP {\mathfrak{k}}
\newcommand \cGP {\mathfrak{c}}

Complex GPs are covered for instance in \cite{miller1969complex}. %
A complex GP is a complex process, where the real and imaginary part
considered jointly are a real GP.
We consider a zero-mean complex-valued random process $\xi$ on
$\Sset$, with covariance function~$\kGP$ and pseudo-covariance
function~$\cGP$:
\begin{align}
  \Esp\left( \xi(s) \xi(s_0)^* \right) & = \kGP(s,s_0), \\
  \Esp \left( \xi(s) \xi(s_0) \right) & = \cGP(s,s_0).
\end{align}

Relying on the mapping~$\Acal$, we can work in a real-valued setting,
i.e., with a real-valued GP $\tilde{\xi}$ indexed
on~$\tilde{\Sset}$. %
In the real-valued case, it is well-known that the conditional mean of
a GP is identical to the minimum-norm interpolant in the RKHS
associated to its covariance function. %
Hence, using~$\Acal$, the conditional mean of a complex GP~$\xi$ is
also identical to a minimum-norm interpolant, but this time in a
complex/real RKHS, the complex kernel~$k$ and pseudo-kernel~$c$ of which are
equal to~$\kGP$ and~$\cGP$ respectively (this follows from
Equations~\eqref{eq:complex-kernel}--\eqref{eq:pseudo-kernel}). %
It is given by Equation~\eqref{equ:interpolant:cr} in general, which
simplifies to
\begin{equation}
  \Esp \left( \xi(s) | y  \right) =  \sum_{i=1}^n \gamma_i k(s,s_i),
  \quad \text{with} \   K_n \gamma  = y,
\end{equation}
if the pseudo-covariance is zero (i.e., in the circular case).


\begin{remark} \label{rem:real-imaginary-independent-interpolation}
  A common approach to deal with complex data is to
  use GP interpolation for the real and imaginary part separately
  (see, e.g., \cite{Fuhrlander_2020ab}). %
  This corresponds, using notations from
  Proposition~\ref{prop:repr-eval-func}, to $k_{RI} = k_{IR} = 0$, and
  therefore to a complex GP with covariance $k = k_{RR} + k_{II}$ and
  pseudo-covariance $c = k_{RR} - k_{II}$.
\end{remark}

\begin{remark} \label{rem:GP-regression_circular}
  GP regression with both covariance and pseudo-covariance function
  has also been considered under the name widely linear posterior
  mean. In \cite{picinbono:1996} it is first shown that the posterior
  mean is widely linear \cite{picinbono1995widely}, which leads to
  \begin{equation}
    \Esp \left( \xi(s) | y  \right) = (k_{s,n} - c_{s,n} K_n^{-*}C_n^H)P_n^{-*}y + (c_{s,n} - k_{s,n} K_n^{-1}C_n)P_n^{-1}y^*,
  \end{equation}
  where $P_n = K_n^* - C_n^H K_n^{-1}C_n$ and $P_n^{-*}$ denotes the
  complex conjugate of the inverse of $P_n$. The formulas for the
  circular and non-circular case can also be found in
  \cite{boloix2018complex}.
\end{remark}

\begin{remark} \label{rem:semi-RKHS} %
  In practice, GP models often include a non-zero mean function~$m$,
  usually written as a linear combination
  $m(x) = \sum_{\ell=1}^L \beta_\ell h_\ell(x)$ of known basis functions~$h_\ell$,
  with unknown coefficients~$\beta_\ell$. %
  If the coefficients are estimated by maximum likelihood (as in
  Section~\ref{sec:alg}), the posterior mean of the GP is then equal
  to the interpolant with minimal \emph{semi-norm} in~$G = V + H$,
  where $V = \vect \{ h_1, \ldots h_L \}$ and the semi-norm is
  defined by
  $\left| g \right|_G = \inf_{v \in V} \lVert g - v \rVert_H$.
\end{remark}

\section{Hybrid algorithm}
\label{sec:alg}
\label{sec:hybrid}
We focus from now, unless otherwise specified, on functions satisfying
the property $f^*(s)=f(s^*)$, and we employ the Szegö
kernel~\eqref{equ:szego}, together with the
pseudo-kernel~\eqref{equ:pseudo-szego}, for complex/real interpolation. %
In practice, the convergence of complex/real RKHS interpolation can be
significantly slower than that of rational approximations techniques
(such as AAA or VF) when the function has a few dominant poles~$p_i$,
i.e., poles with small attenuation~$\Re[p_i] \approx 0$. %
In this section, we discuss how complex/real RKHS interpolation with
the Szegö kernel and associated pseudo-kernel can be combined with a
small number of rational basis functions for the approximation of such
frequency response functions.

\subsection{Gaussian process model}
\label{sec:conjPoles}
We propose to use a complex GP model with rational mean
function~$m = \sum_{\ell = 1}^L \beta_\ell h_\ell$
(cf.~Remark~\ref{rem:semi-RKHS}), covariance
function~$\sigma^2 k_\alpha$ and pseudo-covariance
function~$\sigma^2 c_\alpha$, %
where $k_\alpha$ denotes the Szegö kernel~\eqref{equ:szego},
$c_\alpha$ the associated pseudo-kernel~\eqref{equ:pseudo-szego}, %
and $\sigma^2$, $\alpha$, $\beta_1$, \ldots, $\beta_L$ are
real parameters with $\sigma^2 > 0$ and~$\alpha > 0$. %
For the mean function~$m$ we assume a rational function satisfying the
property $m^*(s) = m(s^*)$, of the form
\begin{equation}\label{eq:basis}
  m(s) = \sum_{i=1}^K \left\{ \frac{1}{s - p_i}\, r_i \,+\, \frac{1}{s - p_i^*}\, r_i^* \right\},
\end{equation}
with residues $r_1, \ldots, r_K \in \Cset$ and (stable) complex
conjugate poles $p_1, p_1^*, \ldots, p_K, p_K^* \in \Cset$ such that
$\Re(p_i) < 0$ and~$\Im(p_i) > 0$ for all~$i$. %
This representation is similar to the one used in VF
\cite{gustavsen2006improving, gustavsen1999rational}. %
Equation~\eqref{eq:basis} can be rewritten
as~$m = \sum_{\ell = 1}^L \beta_\ell h_\ell$ with $L = 2K$,
\begin{equation*}
  \beta_\ell = \begin{cases}
    \Re(r_i) & \text{if } \ell = 2i-1,\\
    \Im(r_i) & \text{if } \ell = 2i,
  \end{cases}
  \quad \text{and} \quad
  h_\ell(s) = \begin{cases}
    \frac{1}{s - p_i} + \frac{1}{s - p_i^*} & \text{if } \ell = 2i-1,\\
    \frac{\i}{s - p_i} - \frac{\i}{s - p_i^*} & \text{if } \ell = 2i.
  \end{cases}
\end{equation*}
Note that $m$ is an element of $H^2_\mathrm{sym}(\Gamma_{\alpha'})$
with $\alpha' = \min_{1\le i \le K} \left| \Re(p_i) \right| + \epsilon, \epsilon > 0$. %
For simplicity we only consider complex conjugate poles in~\eqref{eq:basis}, but
real poles could be included as well, as in VF. %
In the context of the present work, we typically consider a small
number~$K$ of pole pairs ($K \le K_{\mathrm{max}} = \min \left( 5, \lfloor n/4 \rfloor \right)$ in the examples). In the Supplementary Material we investigate different choices of $K_{\mathrm{max}}$ for a specific example and observe that it does not have a large influence.

For a given number~$K$ of pole pairs, we select the hyper-parameters
$\sigma^2$, $\alpha$, $\mathbf p = \left( p_1, \ldots, p_K \right)$
and $\mathbf r = \left( r_1, \ldots, r_K \right)$ by maximization of a
penalized log-likelihood function, where the penalty stems from a vague
log-normal prior on~$\alpha$; %
see Appendix~\ref{sec:penalized-loglik} for details. %
An original procedure for the selection of an appropriate number~$K$
of pole pairs will be presented in the next section.

\begin{remark}
  Note that we do not include a constant basis function, as is
  usually done in Gaussian process modeling, to ensure that the
  interpolant satisfies the desired property (namely, goes to zero)
  when $\omega \to \pm\infty$.
\end{remark}

\begin{remark}
  \label{rmk:mean-to-covariance}
  Assuming that the coefficients $\beta_\ell$ follow a zero-mean Gaussian distribution:
  $\beta_\ell \sim \mathcal{N}(0,\sigma_\ell^2)$, $1 \le \ell \le L$, we could also integrate
  the contribution of the rational mean function $m = \sum_{\ell = 1}^L \beta_\ell h_\ell$
  directly into the covariance and pseudo-covariance functions as
  \begin{equation}
    k_m(s,s_0) = \sum_{\ell=1}^L \sigma_\ell^2\, h_\ell(s)\, h_\ell(s_0)^*,
    \qquad
    c_m(s,s_0) = \sum_{\ell=1}^L \sigma_\ell^2\, h_\ell(s)\, h_\ell(s_0).
  \end{equation}
  This approach has been pursued, e.g., in \cite{hallemans2022frf} and allows the
  uncertainty about the coefficients~$\beta_\ell$ to be reflected in the uncertainty
  estimates of the GP.
  
  We do not pursue this idea further in this article, since our focus is
  on interpolation rather than uncertainty quantification.
\end{remark}

\subsection{Adaptive pole selection}
\label{sec:adap_proc}

Selecting a suitable number~$K$ of pole pairs to be included in the
mean function~\eqref{eq:basis} is a crucial step to ensure good
accuracy of the proposed hybrid method. %
In this section we propose a model selection procedure to select this
number automatically, in a data-driven manner. %
While this procedure relies on the well-established idea of
(leave-one-out) cross-validation, it contains an original ingredient
in the form an ``instability penality'', which will be described
below.

First we build $K_{\max} + 1$ interpolants~$f^{(K)}_n$, where the
superscript $K$ indicates the number of pole pairs, ranging from~$0$
(zero-mean Gaussian process model) to~$K_{\max}$. %
Following standard VF practice \cite{gustavsen1999rational}, we begin
with the maximum number of poles, $K = K_{\max}$, using an equidistant
distribution of poles close to the frequency axis as a starting point
for optimization. %
The other interpolants are then constructed iteratively, going
backwards: at each step optimization is initialized using~$K$ of
the~${K + 1}$ poles selected at the previous step, %
by removing the pole that leads to the smallest decrease of the (penalized) log-likelihood function. %

Model selection is then based on leave-one-out (LOO) cross-validation,
i.e., on the error indicators
\begin{equation}
  \epsilon^K_\LOO
  = \frac 1 n \sum_{i=1}^n \left| f(\i \omega_i)
    - \hat f^{(K)}_{n-1,i} (\i\omega_i) \right|^2,~~K=0,1,\ldots,K_{\max},
  \label{eq:LOO}
\end{equation} 
where $\hat f^{(K)}_{n-1,i}$ denotes a model constructed without the
$i$-th data point. %
Keeping the poles and kernel hyper-parameters fixed, when removing points, makes it possible to reduce the computational effort, but
was found to introduce an undesired preference for models with a
larger number of poles. %
Hence, we employ the LOO criterion with re-tuning, using the poles and
hyper-parameters of ~$f_n^{(K)}$ as an initial guess when
constructing~$\hat f^{(K)}_{n-1, i}$, $1 \le i \le n$.

\begin{figure}
  \begin{flushleft}
    \begin{tikzpicture}[inner frame xsep=0pt,background rectangle/.style={fill=gray!10}, show background rectangle, framed]
      \begin{axis}[grid, xlabel={$\omega$ ($\mathrm s^{-1}$)},
        width=.33\linewidth, height=.3\linewidth,  ylabel near ticks,  xmin=4.5e3, xmax=4.56e3, ymin=-.17e-7, ymax=.17e-7]
        \addplot[TUDa-1b, thick, no marks] table[x=x, y=loo_imag, col sep = comma]{images/model_selection_issue_approximationFine.csv};
        \addplot[TUDa-9b, thick, no marks] table[x=x, y=loo_real, col sep = comma]{images/model_selection_issue_approximationFine.csv};
        \addplot[TUDa-7b,  densely dashed,thick, no marks] table[x=x, y=ref_imag, col sep = comma]{images/model_selection_issue_approximations.csv};
        \addplot[TUDa-4c,  densely dashed,thick, no marks] table[x=x, y=ref_real, col sep = comma]{images/model_selection_issue_approximations.csv};
        \addplot[only marks, mark=*, mark size =1.5pt, forget plot]table[x=x, y=y_real, col sep = comma]{images/model_selection_issue_trainingdata.csv};
        \addplot[only marks, mark=*, mark size =1.5pt,forget plot]table[x=x, y=y_imag, col sep = comma]{images/model_selection_issue_trainingdata.csv};
      \end{axis}
    \end{tikzpicture}\hspace{-1em}
    \begin{tikzpicture}
      \begin{axis}[grid,xlabel={$\omega$ ($\mathrm s^{-1}$)},
        width=.48\linewidth, height=.33\linewidth,  ylabel near ticks,legend style = {font=\small, at={(1,1)},anchor=north east}, legend pos = outer north east, reverse legend, xmin=4500, xmax=5000, ymin=-.3e-7, ymax=.5e-7]
        \addplot[TUDa-1b, thick, no marks] table[x=x, y=loo_imag, col sep = comma]{images/model_selection_issue_approximationFine.csv};
        \addplot[TUDa-9b, thick, no marks] table[x=x, y=loo_real, col sep = comma]{images/model_selection_issue_approximationFine.csv};
        \addplot[TUDa-7b, densely dashed,thick, no marks] table[x=x, y=ref_imag, col sep = comma]{images/model_selection_issue_approximations.csv};
        \addplot[TUDa-4c, densely dashed,thick, no marks] table[x=x, y=ref_real, col sep = comma]{images/model_selection_issue_approximations.csv};
        \addplot[only marks, mark=*, mark size = 1pt,forget plot]table[x=x, y=y_real, col sep = comma]{images/model_selection_issue_trainingdata.csv};
        \addplot[only marks, mark=*, mark size = 1pt,forget plot]table[x=x, y=y_imag, col sep = comma]{images/model_selection_issue_trainingdata.csv};
        \legend{$\Im[f^{(5)}_n(\i\omega)]$, $\Re[f^{(5)}_n(\i\omega)]$,  $\Im[f(\i\omega)]$, $\Re[f(\i\omega)]$}
        \coordinate (spypoint) at (axis cs:4520,0);
      \end{axis}	
      \node[pin={[pin distance=.06\linewidth]200:{%
        }},draw,circle,minimum size=1cm] at (spypoint) {};
    \end{tikzpicture}
  \end{flushleft}\vspace{1.5em}
  \begin{flushleft}
    \begin{tikzpicture}[inner frame xsep=0pt,background rectangle/.style={fill=gray!10}, show background rectangle, framed]
      \begin{axis}[grid, xlabel={$\omega$ ($\mathrm s^{-1}$)},
        width=.33\linewidth, height=.3\linewidth,  ylabel near ticks,  xmin=4500, xmax=4.56e3, ymin=-.17e-7, ymax=.17e-7]
        \pgfplotsinvokeforeach {3,4,6,7, 9,10, 12,13,15,16,18,19, 21,22,24,25,27,28, 30, 31, 33,34,36,37,39,40,42,43,45,46,48,49, 51,52,54,55,57,58, 60,61,63,64}{
          \addplot+[no marks, dashed] table [x =x, y index=#1, col sep = comma] {images/model_selection_issue_loo_predictions.csv};
        }
        \addplot[only marks,  mark=*, mark size =1.5pt, forget plot]table[x=x, y=y_real, col sep = comma]{images/model_selection_issue_trainingdata.csv};
        \addplot[only marks, mark=*, mark size =1.5pt,  forget plot]table[x=x, y=y_imag, col sep = comma]{images/model_selection_issue_trainingdata.csv};
      \end{axis}	
    \end{tikzpicture}\hspace{-1em}
    \begin{tikzpicture}
      \begin{axis}[grid, xlabel={$\omega$ ($\mathrm s^{-1}$)},
        width=.48\linewidth, height=.33\linewidth,  ylabel near ticks, legend style = {font=\small, at={(1.02,1.1)},anchor=north west},  xmin=4.5e3, xmax=5.0e3, ymin=-.3e-7, ymax=.5e-7]
        \pgfplotsinvokeforeach {3,4,6,7, 9,10, 12,13,15,16,18,19, 21,22,24,25,27,28, 30, 31, 33,34,36,37,39,40,42,43,45,46,48,49, 51,52,54,55,57,58, 60,61,63,64}{
          \addplot+[no marks, dashed, thick] table [x =x, y index=#1, col sep = comma] {images/model_selection_issue_loo_predictions.csv};
        }
        \addplot[only marks, mark=*, mark size =1pt,  forget plot]table[x=x, y=y_real, col sep = comma]{images/model_selection_issue_trainingdata.csv};
        \addplot[only marks, mark=*, mark size =1pt,  forget plot]table[x=x, y=y_imag, col sep = comma]{images/model_selection_issue_trainingdata.csv};
        \legend{ $\Re[\hat f^{(5)}_{n-1,1}(\omega)]$,  $\Im[\hat f^{(5)}_{n-1,1}(\omega)]$,  $\Re[\hat f^{(5)}_{n-1,2}(\omega)]$,  $\Im[\hat f^{(5)}_{n-1,2}(\omega)]$, {\Large $\vdots$}}
        \coordinate (spypoint) at (axis cs:4520,0);
      \end{axis}	
      \node[pin={[pin distance=.06\linewidth]200:{%
        }},draw,circle,minimum size=1cm] at (spypoint) {};
    \end{tikzpicture}
  \end{flushleft}
  \caption{Top: Dashed lines show the function to approximate. Black
    dots indicate the training data. Solid lines represent a
    \textit{bad} approximation model which, however, is selected by
    the LOO criterion. %
    Zoomed plot (gray background) highlights the influence of a
    wrongly identified pole. %
    Bottom: Leave-on-out predictions, which show strong local
    variations between \SI{4500}{\per\second} and
    \SI{4520}{\per\second}. %
    However, these variations do not significantly affect the values
    at the respective training points.}
  \label{fig:loo_selection_issue}
\end{figure}
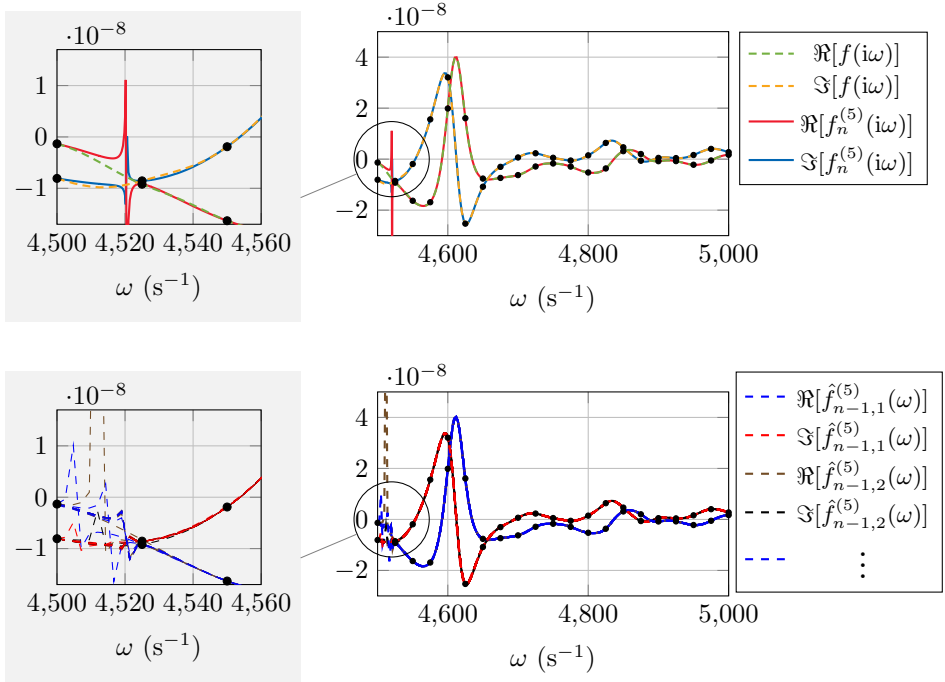

Furthermore, we introduce an additional penalty term, which also takes
\textit{global} model variations into account. %
This approach can be motivated by the example illustrated in
Figure~\ref{fig:loo_selection_issue} (top). %
The corresponding vibro-acoustic benchmark model will be described in
Section~\ref{sec:numerics}, however, here we simply consider the
approximation of the dashed function, based on interpolation of the
training points (black dots), as a general example. %
At the top, it can be observed that the LOO criterion \eqref{eq:LOO}
leads to the selection of a model (solid lines)
$\hat f_n^{(5)}$ which
wrongly identifies a pole at $\approx \SI{4520}{\per \second}$. %
However, this effect is rather local, it mainly takes place between
two training points (illustrated by black dots). %
At the bottom, we show the models
$\hat f^{(K)}_{n-1,i}(\omega),~~i = 1, \ldots, n$, which show strong
variations close to $\approx \SI{4510}{\per \second}$ but rather small
errors at the training points $\omega_i$. %
To take this into account, we introduce an instability penalty term,
which leads to the criterion
\begin{equation}
  \epsilon^K_{\LOO, \mathrm{stab}} = \epsilon^K_\LOO
  + \lambda\, \frac 1 n \frac 1M \sum_{i=1}^n \sum_{j=1}^M \left| f^{(K)}_n (\i\hat \omega_j)
    - \hat f^{(K)}_{n-1,i} (\i\hat \omega_j) \right|^2,
  \label{eq:LOOstab}
\end{equation}
where $\{\hat \omega_j\}_{j=1}^M$ denotes a fine grid on $\Omega$
(more precisely, an equidistant grid with $M = 10n + 1$ points). %
The weighting factor $\lambda$ is chosen as
\begin{equation}
  \lambda = 0.2\, \frac{\epsilon^0_\LOO}{\frac 1 n \frac 1M \sum_{i=1}^n \sum_{j=1}^M \left| f^{(0)}_n (\i\hat \omega_j)
      - \hat f^{(0)}_{n-1,i} (\i\hat \omega_j) \right|^2},
\end{equation}
i.e., $0.2$ after normalizing both terms w.r.t.\ the respective values
of the purely kernel-based interpolation model. %
To our knowledge, this approach for model selection has not been
considered before, although it is related to the continuously-defined
LOO error \cite{jin2002sequential,kim2009construction,
  fuhg2021state}. %
The continuously-defined LOO error was employed for sequential
sampling, while we propose to use it to construct an
instability penalty for model selection. %
Stability selection \cite{liu2020surrogate, meinshausen2010stability}
is another related approach, which is also based on resampling of the
data, but usually employed for variable selection.

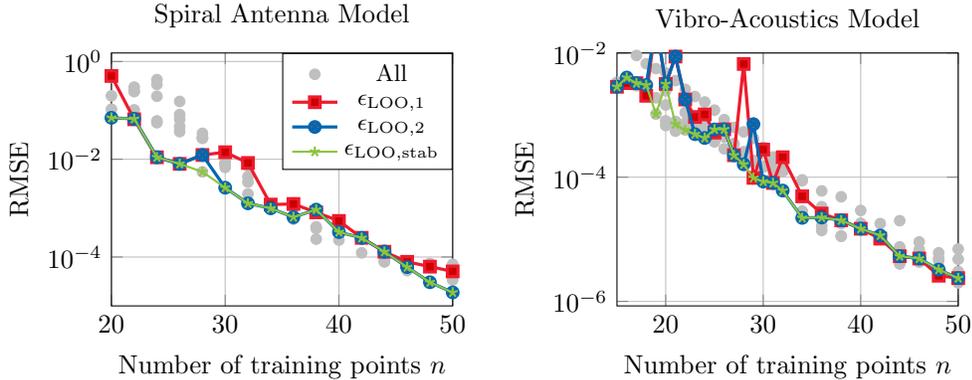
\begin{figure}
  \begin{tikzpicture}
    \begin{semilogyaxis}[grid, xlabel={Number of training points $n$}, ylabel={RMSE}, width=.47\textwidth, height=.38\textwidth, legend style = {font=\small, at={(1,1)},anchor=north east},  xmin=20, xmax=50, clip mode=individual, title={Spiral Antenna Model}, ymin=1e-5]
      \addplot[gray!50, only marks] table[x=N, y=e1, col sep = comma]{images/AllErrors_Spiral.csv};
      \addplot[gray!50, only marks, forget plot] table[x=N, y=e2, col sep = comma]{images/AllErrors_Spiral.csv};
      \addplot[gray!50, only marks, forget plot] table[x=N, y=e3, col sep = comma]{images/AllErrors_Spiral.csv};
      \addplot[gray!50, only marks, forget plot] table[x=N, y=e4, col sep = comma]{images/AllErrors_Spiral.csv};
      \addplot[gray!50, only marks, forget plot] table[x=N, y=e5, col sep = comma]{images/AllErrors_Spiral.csv};
      \addplot[gray!50, only marks, forget plot] table[x=N, y=e6, col sep = comma]{images/AllErrors_Spiral.csv};
      \addplot+[TUDa-9b, very thick] table[x=N, y=LooCheap, col sep = comma]{images/ModelSelectionConvSpiral.csv};
      \addplot+[TUDa-1b, very thick] table[x=N, y=LooExp, col sep = comma]{images/ModelSelectionConvSpiral.csv};
      \addplot+[TUDa-4b, thick] table[x=N, y=CombExp, col sep = comma]{images/ModelSelectionConvSpiral.csv};
      \legend{All, $\epsilon_{\LOO,1}$,$\epsilon_{\LOO,2}$, $\epsilon_{\LOO, \mathrm{stab}}$}
    \end{semilogyaxis}
  \end{tikzpicture}\hfill
  \begin{tikzpicture}
    \begin{semilogyaxis}[grid, xlabel={Number of training points $n$},
      ylabel={RMSE}, width=.47\textwidth, height=.38\textwidth,
      ymax=1e-2, xmin=15, xmax=50, clip mode=individual,
      title={Vibro-Acoustics Model}]
      \addplot[gray!50, only marks] table[x=N, y=e1, col sep = comma]{images/AllErrors_VibroAcoustics.csv};
      \addplot[gray!50, only marks, forget plot] table[x=N, y=e2, col sep = comma]{images/AllErrors_VibroAcoustics.csv};
      \addplot[gray!50, only marks, forget plot] table[x=N, y=e3, col sep = comma]{images/AllErrors_VibroAcoustics.csv};\\
      \addplot[gray!50, only marks, forget plot] table[x=N, y=e4, col sep = comma]{images/AllErrors_VibroAcoustics.csv};
      \addplot[gray!50, only marks, forget plot] table[x=N, y=e5, col sep = comma]{images/AllErrors_VibroAcoustics.csv};
      \addplot[gray!50, only marks, forget plot] table[x=N, y=e6, col sep = comma]{images/AllErrors_VibroAcoustics.csv};
      \addplot+[TUDa-9b, very thick] table[x=N, y=LooCheap, col sep = comma]{images/ModelSelectionConvVibroAcoustics.csv};
      \addplot+[TUDa-1b, very thick] table[x=N, y=LooExp, col sep = comma]{images/ModelSelectionConvVibroAcoustics.csv};
      \addplot+[TUDa-4b, thick] table[x=N, y=CombExp, col sep = comma]{images/ModelSelectionConvVibroAcoustics.csv};
    \end{semilogyaxis}
  \end{tikzpicture}
  \caption{Comparison of different model selection criteria for two
    benchmark problems. %
    $\epsilon_{\LOO,1}$ and $\epsilon_{\LOO,2}$ denote
    the leave-on-out residual without and with retuning of
    hyper-parameters, respectively. %
    The stabilized criterion $\epsilon_{\LOO, \mathrm{stab}}$ (with
    retuning) defined in \eqref{eq:LOOstab} gives the best results.}
  \label{fig:conv_error_indicators}
\end{figure}

Employing the stabilized criterion \eqref{eq:LOOstab} for model
selection gives satisfactory results for the benchmark examples
considered in this work. %
For illustration, we consider the convergence studies for two models,
which will be described in Section~\ref{sec:numerics}. %
Figure~\ref{fig:conv_error_indicators} shows the root-mean-square-errors
(RMSEs) of the available models with gray dots and the accuracy of the
selected models by the different criterions. %
It can be observed that the stabilized criterion
$\epsilon^K_{\LOO, \mathrm{stab}}$ gives the best results, while LOO
residuals with retuning is superior to the approach without
retuning.

\begin{remark}\label{rmk:hallemans}%
  The combination of kernel methods with a small number of rational
  basis functions has also been considered in \cite{hallemans2022frf}
  for data-driven modeling of frequency response functions. %
  Therein, the authors employ first order stable spline kernels, which
  encode stability, causality and smoothness and add a rational basis
  for capturing the resonsant poles of the transfer function. %
  A prior is formed over the impulse responses linked to the resonant
  poles, which allows to derive additional kernels (one for each
  resonant pole) via the Fourier transform. %

  Our approach proceeds in a similar way, as our VF-inspired rational
  basis could also be transformed into additional kernels through a
  prior over $\beta$, see also Remark~\ref{rmk:mean-to-covariance}. %
  Contrary to \cite{hallemans2022frf}, we provide a complete background on the RKHS concepts of complex/real interpolation. Further differences can be found in the choice of kernels and the way the mean functions are constructed. The mean function is obtained in \cite{hallemans2022frf} with the local rational method, which first constructs rational approximations of variable order in local frequency bands. Then, in each local band, the dominant pole is selected, all poles are clustered and from each cluster the pole with the lowest variance is selected to be included into the mean function. This approach is very flexible, but also complex with several algorithmic steps that need to be coupled. Instead, our approach employs a global rational approximation and selects the number of poles based on statistical principles with a new model selection criterion.
  It should be noted that \cite{hallemans2022frf} is additionally targeting uncertainty
  quantification for the data-driven modeling procedure.
\end{remark}

\section{Numerical results}
\label{sec:numerics}
We apply the presented approximation techniques to a number of
benchmark functions from different fields. %
We always employ $n$ training points $(\omega_i,f(\i\omega_i)),$ where
the $\omega_i$ are equidistant frequency points in
$[\omega_{\min},\omega_{\max}]$, for simplicity. %
The accuracy of different approximations is then quantified in terms
of the root-mean-square error (RMSE), which is evaluated on a refined
equidistant grid with $201$ points for all numerical examples.

In the following we give a few details on the implementation. %
For AAA \cite{nakatsukasa2018aaa}, we rely on the implementation of
the \texttt{chebfun} toolbox \cite{driscoll2014chebfun}. %
For VF, we employ the \texttt{VectFit3} toolbox
\cite{gustavsen1999rational,gustavsen2006improving,deschrijver2008macromodeling},
where we use complex equidistant starting poles distributed according
to the general recommendation, and always run 30 iterations. %
We apply the ``relaxed non-triviality constraint''
\cite{gustavsen2006improving}, include the constant but not the linear
term, and enforce stable poles. %
The number of complex starting pole pairs is set to the maximum number
of $2\lfloor \frac {n-1}2\rfloor$, which leads to the best results for
the smooth test functions considered. %
For kernel interpolation we consider a separate interpolation of the
real and imaginary part with the squared exponential kernel (SE) and
complex/real interpolation with the Szegö kernel. %
The latter is also considered in combination with an adaptive rational
basis (Sz.-Rat.)\ as described in Section~\ref{sec:hybrid}. %
The implementation is done in \texttt{Matlab}, based on the
\texttt{STK} toolbox \cite{bect:2023:stk}  and the code to reproduce all results is publicly available \cite{georg_2024_12601365}. %
We employ the mapping $\mathcal A$ defined in
\eqref{eq:mappingNonIntrusive} for the complex/real RKHS
interpolation, which allows to realize the implementation based on
real RKHS interpolation on an augmented input space
$\Omega \times\{0,1\}$. %
Note that this approach could be employed with any toolbox for real
RKHS interpolation that provides the option to specify custom kernel
functions. %
The tuning of the hyper-parameters and poles based on the likelihood
function (see Section~\ref{sec:alg}) is carried out using
\texttt{fmincon} in Matlab, i.e., gradient-based optimization (more
precisely an interior point algorithm), which we combine with a
multistart procedure; see Appendix~\ref{sec:details-param-optim} for more details.

\begin{remark}
  By investigating the shape of the likelihood function for a number
  of benchmark problems, we have found that the logarithmic
  reparameterization, discussed in \cite{basak:2022:numerical} for
  instance, is not beneficial for the parameter $\alpha$. %
  Hence, it is only applied to the scaling parameter~$\sigma$.
\end{remark}

\subsection{Electric circuit (high order rational function)}

We consider in the following a parallel connection of $N$ underdamped
series RLC circuits, as illustrated on the left side in
Figure~\ref{fig:Electric_circuit}. %
The admittance is given as
\begin{equation}
  Y(s) = \sum_{i=1}^N \frac
  {s}{s^2 L_i +s R_i + C_i^{-1}}=\sum_{i=1}^N \frac {c_i}{s-a_i}+\frac
  {c_i^*}{s -a_i^*},\label{eq:admittance}
\end{equation}
where $\Re[a_i] = -\frac{R_i}{2L_i}$ (an explicit representation of
the poles $a_i$ and residues $c_i$ is given in the Supplementary
Material) and we consider the frequency range
$[\SI{10}{kHz}, \SI{25}{kHz}]$. %
First, we assume $N_1=1000$ random series RLC elements, where
$C_i\sim \mathcal U(1,20)\,\SI{}{\micro\farad}$ and
$L_i\sim\mathcal U(0.1,2)\,\SI{}{\milli\henry}$, %
and we assume the resistance $R_i$ to be roughly proportional to the
inductance, with random variations of $\pm 20\%$:
$R_i=L_i(1+\Delta)\, \SI{}{\ohm}(\SI{}{\milli\henry})^{-1}$, where
$\Delta \sim \mathcal U(-0.2, 0.2)$.

\begin{figure}
  \begin{tikzpicture}
    \begin{axis}[grid,
      xlabel={$\omega$}, 
      width=.355\textwidth, height=.35\textwidth, ylabel near ticks,
      legend style = {font=\small}, legend pos = south west, title = {
        $\operatorname{Re}[Y(\i\omega)]$}, xmin=1e4, xmax=2.5e4]
      \addplot[TUDa-9b, very thick, no marks] table[x=x, y=y_ref_real,
      col sep = comma]{images/illustration_CircuitDomPoles.csv};
      \addplot[black, very thick, no marks, densely dashed] table[x=x,
      y=y_ref_real, col sep = comma]{images/illustration_Circuit.csv};
    \end{axis}
  \end{tikzpicture}
  \begin{tikzpicture}
    \begin{axis}[grid,
      xlabel={$\omega$}, 
      width=.355\textwidth, height=.35\textwidth, ylabel near ticks,
      legend style = {font=\small}, legend pos = north east, title = {
        $\operatorname{Im}[Y(\i\omega)]$}, xmin=1e4, xmax=2.5e4,
      reverse legend]
      \addplot[TUDa-9b, very thick, no marks] table[x=x, y=y_ref_imag,
      col sep = comma]{images/illustration_CircuitDomPoles.csv};
      \addplot[black, very thick, no marks, densely dashed] table[x=x,
      y=y_ref_imag, col sep = comma]{images/illustration_Circuit.csv};
      \legend{$Y_2(\i\omega)$, $Y_1(\i\omega)$}
    \end{axis}
  \end{tikzpicture}
  \begin{tikzpicture}
    \begin{axis}[grid,
      xlabel={$\omega$}, 
      width=.355\textwidth, height=.35\textwidth, ylabel near ticks,
      legend style = {font=\small}, legend pos = north east, title = {
        $|Y(\i\omega)|$} , xmin=1e4, xmax=2.5e4]
      \addplot[TUDa-9b, very thick, no marks] table[x=x, y=y_ref_abs,
      col sep = comma]{images/illustration_CircuitDomPoles.csv};
      \addplot[black, very thick, densely dashed, no marks] table[x=x,
      y=y_ref_abs, col sep = comma]{images/illustration_Circuit.csv};
    \end{axis}
  \end{tikzpicture}
  \caption{Complex admittances $Y_1$ and $Y_2$ of the electric
    circuits versus frequency for a particular random parameter
    realization and $N_1=1000$ and $N_2=1002$, respectively.}
  \label{fig:fun_circuit}
\end{figure}
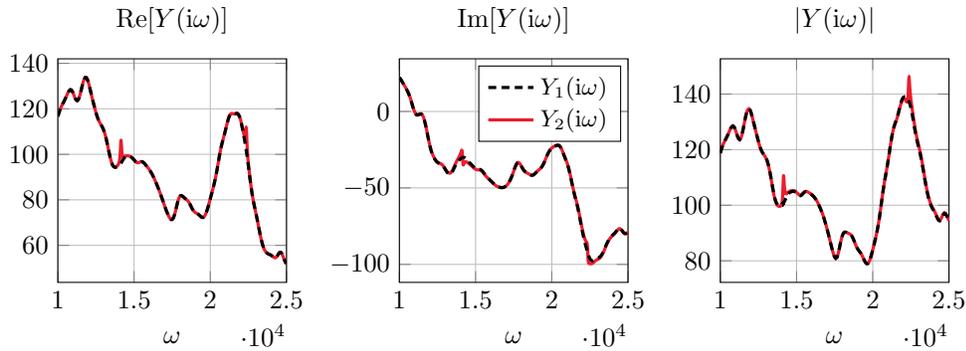

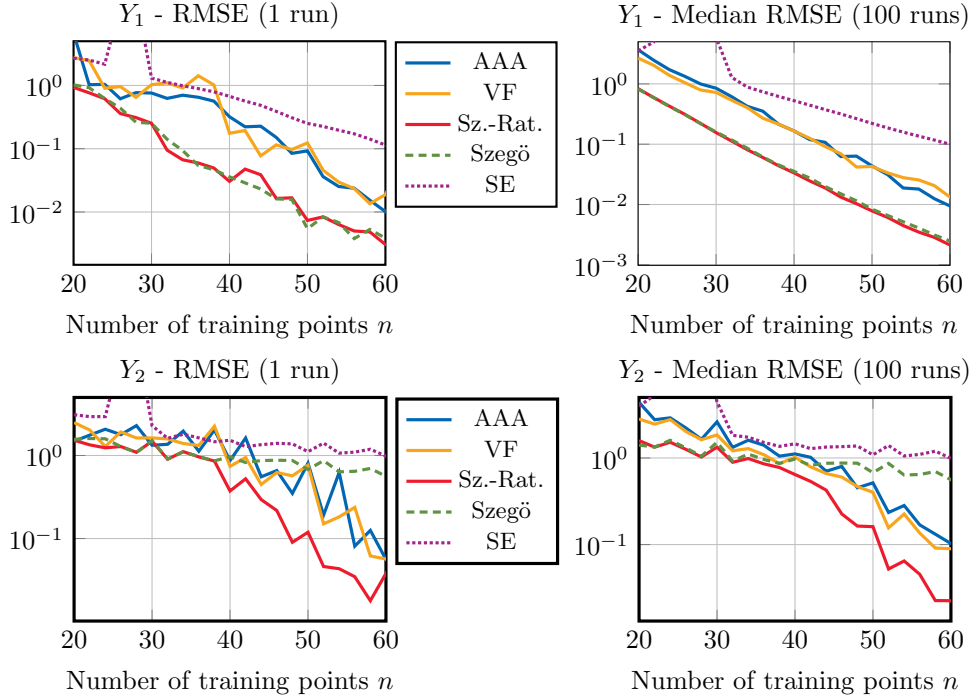
\begin{figure}
  \begin{tikzpicture}
    \begin{semilogyaxis}[grid, xlabel={Number of training points $n$},
      title={$Y_1$ - RMSE (1 run)}, width=.44\textwidth,
      height=.35\textwidth, legend pos = outer north east, legend
      style = {font=\small}, no marks, thick, ymax=5, xmin=20, ytick =
      {1e0, 1e-1, 1e-2, 1e-3, 1e-4}, every axis plot/.append
      style={very thick}, title style = {yshift=-.5em}, xmax=60]
      \addplot[TUDa-1b] table[x=N, y=RMSE, col sep = comma]{images/Circuit_AAA.csv};
      \addplot[TUDa-7b] table[x=N, y=RMSE, col sep = comma]{images/Circuit_VF.csv};
      \addplot[TUDa-9b, very thick] table[x=N, y=RMSE, col sep = comma]{images/Circuit_Adap.csv};	
      \addplot[TUDa-4d, densely dashed] table[x=N, y=RMSE, col sep = comma]{images/Circuit_Szego.csv};
      \addplot[TUDa-10b, densely dotted] table[x=N, y=RMSE, col sep = comma]{images/Circuit_GaussSep.csv};
      \legend{AAA, VF, Sz.-Rat.,Szegö, SE}
    \end{semilogyaxis}
  \end{tikzpicture}
  \hfill
  \begin{tikzpicture}
    \begin{semilogyaxis}[grid, xlabel={Number of training points $n$},
      title={$Y_1$ - Median RMSE (100 runs)}, width=.44\textwidth,
      height=.35\textwidth, legend pos = north east, legend style =
      {font=\small}, no marks, ymax=5, xmin=20, ytick = {1e0, 1e-1,
        1e-2, 1e-3, 1e-4}, every axis plot/.append style={very thick},
      title style = {yshift=-.5em}, xmax=60]
      \addplot[TUDa-1b] table[x=N, y=RMSE, col sep = comma]{images/Circuit100runs_AAA.csv};
      \addplot[TUDa-7b] table[x=N, y=RMSE, col sep = comma]{images/Circuit100runs_VF.csv};
      \addplot[TUDa-9b,very thick] table[x=N, y=RMSE, col sep = comma]{images/Circuit100runs_Adap.csv};
      \addplot[TUDa-4d, densely dashed] table[x=N, y=RMSE, col sep = comma]{images/Circuit100runs_Szego.csv};
      \addplot[TUDa-10b, densely dotted] table[x=N, y=RMSE, col sep = comma]{images/Circuit100runs_GaussSep.csv};
    \end{semilogyaxis}
  \end{tikzpicture}\\
  \begin{tikzpicture}
    \begin{semilogyaxis}[grid, xlabel={Number of training points $n$},
      title={$Y_2$ - RMSE (1 run)}, width=.44\textwidth,
      height=.35\textwidth, legend pos = outer north east, legend
      style = {font=\small}, no marks, very thick, ymax=5, xmin=20,
      ytick = {1e0, 1e-1, 1e-2, 1e-3, 1e-4},every axis plot/.append
      style={very thick}, title style = {yshift=-.5em}, xmax=60]
      \addplot[TUDa-1b] table[x=N, y=RMSE, col sep = comma]{images/CircuitDomPoles_AAA.csv};
      \addplot[TUDa-7b] table[x=N, y=RMSE, col sep = comma]{images/CircuitDomPoles_VF.csv};
      \addplot[TUDa-9b, very thick] table[x=N, y=RMSE, col sep = comma]{images/CircuitDomPoles_Adap.csv};
      \addplot[TUDa-4d, densely dashed] table[x=N, y=RMSE, col sep = comma]{images/CircuitDomPoles_Szego.csv};
      \addplot[TUDa-10b,densely dotted] table[x=N, y=RMSE, col sep = comma]{images/CircuitDomPoles_GaussSep.csv};
      \legend{AAA, VF, Sz.-Rat., Szegö, SE}
    \end{semilogyaxis}
  \end{tikzpicture}
  \hfill
  \begin{tikzpicture}
    \begin{semilogyaxis}[grid, xlabel={Number of training points $n$},
      title={$Y_2$ - Median RMSE (100 runs)}, width=.44\textwidth,
      height=.35\textwidth, legend pos = outer north east, legend
      style = {font=\small}, no marks, very thick, ymax=5, xmin=20,
      ytick = {1e0, 1e-1, 1e-2, 1e-3}, every axis plot/.append
      style={very thick}, title style = {yshift=-.5em}, xmax=60]
      \addplot[TUDa-1b] table[x=N, y=RMSE, col sep = comma]{images/CircuitDomPoles100runs_AAA.csv};
      \addplot[TUDa-7b] table[x=N, y=RMSE, col sep = comma]{images/CircuitDomPoles100runs_VF.csv};
      \addplot[TUDa-9b, very thick] table[x=N, y=RMSE, col sep = comma]{images/CircuitDomPoles100runs_Adap.csv};
      \addplot[TUDa-4d, densely dashed] table[x=N, y=RMSE, col sep = comma]{images/CircuitDomPoles100runs_Szego.csv};
      \addplot[TUDa-10b, densely dotted] table[x=N, y=RMSE, col sep = comma]{images/CircuitDomPoles100runs_GaussSep.csv};
    \end{semilogyaxis}
  \end{tikzpicture}
  \caption{%
    Convergence study for admittances $Y_1$ (top) and $Y_2$
    (bottom). %
    Left: RMSE for one particular realization. %
    Right: Median for 100 random realizations.}
  \label{fig:ConvStudyCircuit}
\end{figure}

Note that for any combination of those parameters, the corresponding
series RLC circuits are underdamped. %
For one particular realization, the distribution of the $2N = 2000$
poles is illustrated in Figure~\ref{fig:Electric_circuit}. %
The corresponding admittance $Y_1(\i\omega)$ is shown in
Figure~\ref{fig:fun_circuit} with dashed black lines. %
We then conduct a convergence study for the particular realization of
the electric circuit, which is shown in
Figure~\ref{fig:ConvStudyCircuit} (top, left). %
We repeat the convergence study for 100 random realizations and depict
the median RMSE at each point in Figure~\ref{fig:ConvStudyCircuit} (top,
right). %
It can be observed that for the considered range of the number of
training points (where $n \le 60 \ll N$) the complex/real Szegö kernel-based interpolation outperforms
AAA and~VF. This can be explained by the fact that $Y$ is not a low order rational function in this case and the nonparametric kernel approach seems to be better suited to cope with the large number of densely located poles. Employing the hybrid algorithm (Sz.-Rat.)\ does not yield an improvement, but leads to similarly good results.

\begin{figure}
  \centering
  \resizebox{.6\textwidth}{!}{
    \large
    \begin{circuitikz}[american,line width=0.25mm, scale=0.9]
      \foreach \x in {1,2,3,4}
      \draw (2*\x,5) to [R=$R_\x$,*-] (2*\x,3)
      to [L=$L_\x$] (2*\x,1.5)
      to [C=$C_\x$, -*] (2*\x,0);
      \draw (12,5) to [R=$R_N$,*-] (12,3)
      to [L=$L_N$] (12,1.5)
      to [C=$C_N$, -*] (12,0);
      \node at (10, 2.5) {\Huge $\ldots$};            
      \draw (2,5) to (12,5);
      \draw (2,0) to (12,0);
      \draw (7,5) to [short, -*] (7,6);
      \draw (7,0) to [short, -*] (7,-1);
    \end{circuitikz}}\hfill
  \begin{tikzpicture}
    \begin{axis}[grid, xlabel={Real part}, ylabel={Imaginary part},
      width=.4\textwidth, height=.35\textwidth, xmin=-830, xmax=100,
      ymin=-3.8e4, ymax=3.8e4, title={Poles}, ylabel near ticks]
      \addplot[only marks, mark = x] table[x=Poles_Re, y=Poles_Im, col sep = comma]{images/Circuit_VF_representation.csv};
      \addplot[only marks, TUDa-9b, mark = x] table[x=Poles_Re, y=Poles_Im, col sep = comma]{images/Circuit_VF_representation_DomPoles.csv};
      \addplot[TUDa-1b, no marks, very thick] coordinates {(0,1e4) (0,2.5e4)};
    \end{axis}
  \end{tikzpicture}
  \caption{%
    Left: Parallel connection of (underdamped) series RLC circuits. %
    Right: Black crosses indicate the distribution of $2 N_1=2000$
    poles of the circuit admittance $Y_1$ in the complex plane. %
    Red crosses indicate the two additional poles considered for the
    circuit admittance $Y_2$ with $2 N_2=2004$ poles. %
    Blue line indicates the considered frequency range.}
  \label{fig:Electric_circuit}
\end{figure}
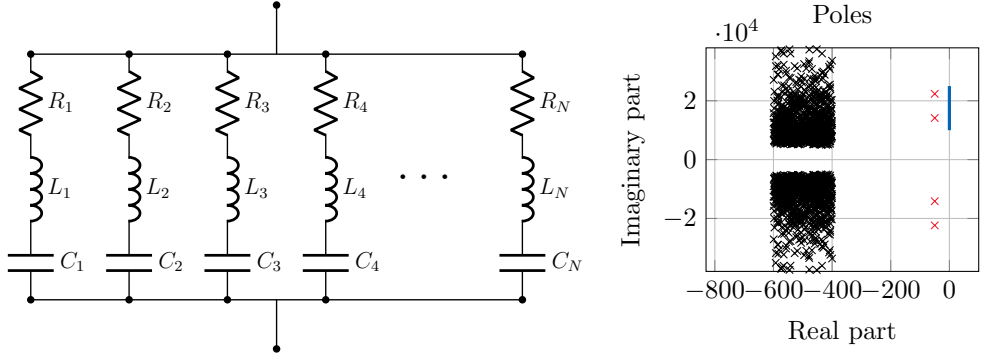

\begin{figure}
	\begin{center}
		\begin{tikzpicture}
		\begin{axis}[grid, width=.85\textwidth, height=.4\textwidth, ylabel near ticks,
		legend style = {font=\small}, legend pos = outer north east,
		xmin=1e4, xmax=2.5e4]
		\addplot[black, very thick, no marks] table[x=x, y=y_ref_abs,
		col sep = comma]{images/illustration_approx_circuit.csv};
		\addplot[TUDa-1b, very thick, no marks] table[x=x, y=AAA,
		col sep = comma]{images/illustration_approx_circuit.csv};
		\addplot[TUDa-7b, dashed, very thick, no marks] table[x=x, y=VF,
		col sep = comma]{images/illustration_approx_circuit.csv};
		\addplot[TUDa-4d, dash dot, very thick, no marks] table[x=x, y=Szego,
		col sep = comma]{images/illustration_approx_circuit.csv};
		\addplot[TUDa-9b, dotted, very thick, no marks] table[x=x, y=Adap,
		col sep = comma]{images/illustration_approx_circuit.csv};
		\legend{$|Y_2|$, AAA, VF, Szegö, Sz.-Rat.}
		\end{axis}
		\end{tikzpicture}
		\begin{tikzpicture}
		\begin{semilogyaxis}[grid,
		xlabel={$\omega$}, 
		width=.85\textwidth, height=.4\textwidth, ylabel near ticks,
		legend style = {font=\small}, legend pos = outer north east,
		xmin=1e4, xmax=2.5e4]
		\addplot[TUDa-1b, very thick, no marks] table[x=x, y=Error_AAA,
		col sep = comma]{images/illustration_approx_circuit.csv};
		\addplot[TUDa-7b, dashed, very thick, no marks] table[x=x, y=Error_VF,
		col sep = comma]{images/illustration_approx_circuit.csv};
		\addplot[TUDa-4d, dashed, very thick, no marks] table[x=x, y=Error_Szego,
		col sep = comma]{images/illustration_approx_circuit.csv};
		\addplot[TUDa-9b, very thick, no marks] table[x=x, y=Error_Adap,
		col sep = comma]{images/illustration_approx_circuit.csv};
		\legend{AAA, VF, Szegö, Sz.-Rat.}
		\end{semilogyaxis}
		\end{tikzpicture}
	\end{center}
	\caption{Magnitude of approximations of complex admittance $Y_2$ (top) and associated errors (bottom) of the electric circuit versus frequency for a particular random parameter realization with different approximation approaches for $n=50$ equidistant training points.}
	\label{fig:fun_circuit_approx}
\end{figure}

In our second experiment, we introduce two additional circuit elements
with a very small damping, i.e., we now consider $N_2 = 1002$ and
\begin{align*}
  C_{1001} = \SI{5}{pF,} && L_{1001} =\SI{1}{mH}, && R_{1001}=\SI{0.1}{\ohm},\\
  C_{1002} = \SI{2}{pF}, && L_{1002} =\SI{1}{mH}, && R_{1002}=\SI{0.1}{\ohm}.
\end{align*}
This leads to two additional poles which are closer to the input
domain, as illustrated by the red crosses in
Figure~\ref{fig:Electric_circuit}. %
The corresponding admittance $Y_2(\i \omega)$ differs very little from
$Y_1(\i\omega)$, except for two sharp peaks, as can be seen in
Figure~\ref{fig:fun_circuit}. %
However,
the accuracy of the respective RKHS interpolation is significantly
affected. %
In particular, at the bottom of Figure~\ref{fig:ConvStudyCircuit}, it
can be observed that the convergence order of Szegö kernel
interpolation is significantly reduced. %
By adding the rational basis we are able to mitigate the impact of the
two dominant poles: it exhibits fast convergence and an improvement
w.r.t.\ AAA and~VF can again be observed. %
This is further illustrated in Fig.~\ref{fig:fun_circuit_approx}, where
the accuracy of the different approximations for the case of $n=50$
training points is shown. %
It can be observed that for all methods the largest approximation
errors occur close to the two sharp peaks, which correspond to the
poles close to the imaginary axis, and that the proposed approach
yields the highest accuracy over the whole frequency range for the
considered test case.%

Next, we comment on the selected values of~$\alpha$ for the
kernel-based approximations. %
For low-order rational functions as considered in
Section~\ref{sec:theory}, and using a higher number of training points,
we show empirically in the supplementary material (see \ref{sec:alpha}) that
the selected values of $\alpha$ are numerically close to the real part
of the poles. %
For the particular test case discussed above, i.e., the random
realization of the electric circuit shown in
Fig.~\ref{fig:fun_circuit}, and using $N=50$ training points %
(which is order of magnitudes below the order of the approximated
rational function), we obtain the following behavior: %
Considering, the pure Szegö-kernel based approximation, i.e., $m(s)=0$,
we obtain $\alpha= 536.99$ and
$\alpha = 353.42$ as selected values of $\alpha$ for $Y_1$
and $Y_2$, which have dominant poles at $\mathbb R[s]\approx-400$ and
$\mathbb R[s]=-50$, respectively. %
As expected, the method chooses a smaller value of $\alpha$ to account
for the extra poles of $Y_2$, however, the values are not yet very
close to the real part of the dominant poles of the approximated
functions. %
Using a non-zero rational mean function, we obtain a larger value of
$\alpha = 414.44$ for $Y_2$, which appears reasonable as
the mean function can partially compensate the impact of the dominant
poles close to the axis.%

\subsection{PDE-based examples}

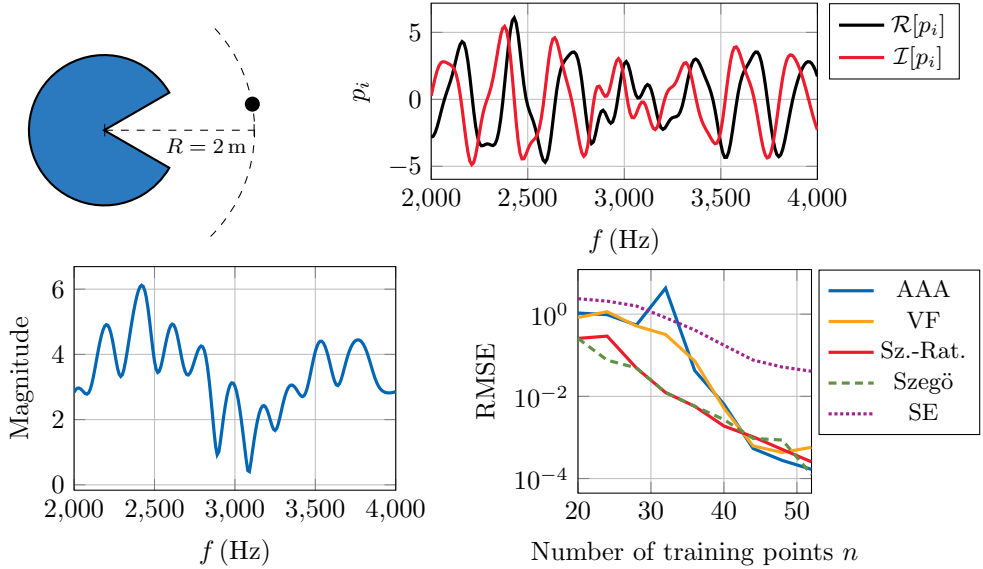
\begin{figure}
  \begin{subfigure}[b]{.3\textwidth}
    \centering
    \begin{tikzpicture}
      \footnotesize
      \draw[thick,fill=TUDa-1b!80]
      (0,0) -- (30:1) arc (30:330:1) -- cycle;
      \fill (1.96961551,0.347296355) circle (1mm);
      \draw[dashed] ({sqrt(2)},{-sqrt(2)}) arc (-45:45:2); 
      \draw[dashed, |-|] (0,0) -- (2,0) node[anchor=north east] {$R=\SI{2}{m}$};
    \end{tikzpicture}
    \vspace{1em}
  \end{subfigure}
  \begin{subfigure}[b]{.67\textwidth}
    \centering
    \begin{tikzpicture}
      \begin{axis}[grid,
        xlabel={$f$\,(Hz)}, 
        width=.77\linewidth, height=.45\linewidth, ylabel near ticks,
        legend style = {font=\small}, legend pos = outer north east,
        xmin=2000, xmax=4000, ylabel = {$p_i$}]
        \addplot[black, very thick, no marks] table[x=x, y=y_ref_real,
        col sep = comma]{images/illustration_PacmanRight.csv};
        \addplot[TUDa-9b, very thick, no marks] table[x=x,
        y=y_ref_imag, col sep =
        comma]{images/illustration_PacmanRight.csv};
        \legend{$\mathcal R[p_i]$, $\mathcal I[p_i]$}
      \end{axis}
    \end{tikzpicture}
  \end{subfigure}
  \begin{tikzpicture}
    \begin{axis}[grid,
      xlabel={$f$\,(Hz)}, 
      width=.45\linewidth, height=.35\linewidth, ylabel near ticks,
      legend style = {font=\small}, legend pos = outer north east,
      xmin=2000, xmax=4000, ylabel = {Magnitude}]
      \addplot[TUDa-1b, very thick, no marks] table[x=x, y=y_ref_abs,
      col sep = comma]{images/illustration_PacmanRight.csv};
    \end{axis}
  \end{tikzpicture}
  \hfill
  \begin{tikzpicture}
    \begin{semilogyaxis}[grid, xlabel={Number of training points $n$},
      ylabel={RMSE}, width=.36\textwidth, height=.35\textwidth, legend
      pos = outer north east, legend style = {font=\small}, xmin = 20,
      xmax=52, every axis plot/.append style={very thick}, no marks]
      \addplot+[TUDa-1b, mark options={fill=TUDa-1b}] table[x=N, y=RMSE, col sep = comma]{images/PacmanRight_AAA.csv};
      \addplot+[TUDa-7b, mark options={fill=TUDa-7b}] table[x=N, y=RMSE, col sep = comma]{images/PacmanRight_VF.csv};
      \addplot+[TUDa-9b, very thick,mark=triangle*, mark options={fill=TUDa-9b}] table[x=N, y=RMSE, col sep = comma]{images/PacmanRight_Adap.csv};
      \addplot+[TUDa-4d, densely dashed, mark options={fill=TUDa-4d}] table[x=N, y=RMSE, col sep = comma]{images/PacmanRight_Szego.csv};
      \addplot+[TUDa-10b, densely dotted, mark options={fill=TUDa-10b}] table[x=N, y=RMSE, col sep = comma]{images/PacmanRight_GaussSep.csv};
      \legend{AAA, VF, Sz.-Rat., Szegö, SE}
    \end{semilogyaxis}
  \end{tikzpicture}
  \caption{%
    Top left: We consider a surface vibration of the PAC-MAN model and
    evaluate the radiated acoustic field $p_i$ at a point (black dot)
    in \SI{2}{m} distance to the center. %
    Top right: Complex frequency response function. %
    Bottom left: Magnitude of frequency response function. %
    Bottom right: Convergence study w.r.t.\ the number of training
    points.}
  \label{fig:Pacman}
\end{figure}

In the following, we investigate a number of PDE-based examples. %
We start with the acoustic Helmholtz equation, in particular, the
PAC-MAN benchmark example, introduced in \cite{ziegelwanger2017pac}
which is also included in the platform for benchmark cases in
computational acoustics from the European Acoustics Association
\cite{hornikx2015platform}. %
The model, shown in Figure~\ref{fig:Pacman}, has the PAC-MAN shape with
an opening angle of $30^\circ$ and radius of $\SI{1}{m}$. %
As in \cite[Section 6.1]{ziegelwanger2017pac}, we consider as
excitation a vibration of the surface of the PAC-MAN with cylindrical
modes and observe the radiated field $u_i$ at a point in \SI{2}{m}
distance at an angle of $10^\circ$. %
As in \cite{hornikx2015platform}, the computation was done based on
the implementation of the analytical solution provided in
\cite{ziegelwanger2017pac} by replacing the python module
\texttt{scipy} by \texttt{mpmath} for the computation of higher order
Bessel functions. %
In particular, we set the truncation order to~300. %
The complex acoustic pressure field phasor $u_i$ of the total
sound-field versus the frequency
$f\in[\SI{2000}{\hertz},\SI{4000}{\hertz}]$ is shown in
Figure~\ref{fig:Pacman} (top, right). %
We then conduct a convergence study w.r.t.\ the number of training
points, which is depicted in Figure~\ref{fig:Pacman} (bottom, left). %
It can be observed that the complex/real Szegö kernel-based interpolation
outperforms the alternative approaches in the range up to about 40
training points. %
Adding the rational mean function does not further improve the
accuracy, but does not harm the accuracy either.

\begin{figure}
  \begin{subfigure}[b]{.4\textwidth}
    \centering
    \includegraphics[width=1.0\textwidth]{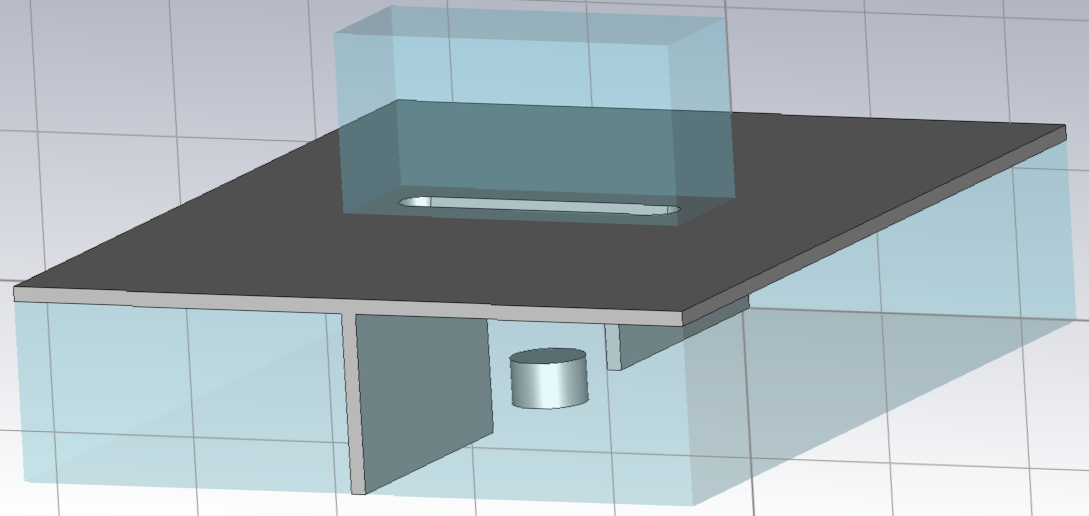}
    \vspace{.5em}
  \end{subfigure}\hfill
  \begin{subfigure}[b]{.59\textwidth}
    \centering
    \begin{tikzpicture}
      \begin{axis}[grid, xlabel={$f$\,(GHz)}, width=.7\linewidth,
        height=.45\linewidth, ylabel near ticks, legend style =
        {font=\small}, legend pos = outer north east, xmin=7, xmax=9,
        every axis plot/.append style={thick}, title style =
        {yshift=-.5em}]
        \addplot[black, thick, no marks] table[x=x, y=y_ref_real, col sep = comma]{images/illustration_WGjunctionS21.csv};
        \addplot[TUDa-9b, thick, no marks] table[x=x, y=y_ref_imag, col sep = comma]{images/illustration_WGjunctionS21.csv};
        \addplot[TUDa-1b, thick, no marks] table[x=x, y=y_ref_real, col sep = comma]{images/illustration_WGjunctionS41.csv};
        \addplot[TUDa-7c, thick, no marks] table[x=x, y=y_ref_imag, col sep = comma]{images/illustration_WGjunctionS41.csv};
        \legend{$\mathcal R[S_{21}]$, $\mathcal I[S_{21}]$, $\mathcal R[S_{41}]$, $\mathcal I[S_{41}]$}
      \end{axis}
    \end{tikzpicture}
  \end{subfigure}
  \begin{subfigure}[b]{.53\textwidth}
    \centering
    \begin{tikzpicture}
      \begin{semilogyaxis}[grid, xlabel={Number of training points
          $n$}, title={$S_{21}$ - RMSE}, width=5cm, height=4cm, legend
        pos = outer north east, legend style = {font=\small}, very
        thick, xmin=5, xmax=15, every axis plot/.append style={very
          thick}, no marks, title style = {yshift=-.5em}]
        \addplot+[TUDa-1b,  mark options = {fill=TUDa-1b}] table[x=N, y=RMSE, col sep = comma]{images/WGjunctionS21_AAA.csv};
        \addplot+[TUDa-7b, mark options = {fill=TUDa-7b}] table[x=N, y=RMSE, col sep = comma]{images/WGjunctionS21_VF.csv};
        \addplot[TUDa-9b, mark=triangle*, mark options={fill=TUDa-9b}] table[x=N, y=RMSE, col sep = comma]{images/WGjunctionS21_Adap.csv};
        \addplot+[TUDa-4d, densely dashed, mark options={fill=TUDa-4d}] table[x=N, y=RMSE, col sep = comma]{images/WGjunctionS21_Szego.csv};
        \addplot+[TUDa-10b, densely dotted, mark options={fill=TUDa-10b}] table[x=N, y=RMSE, col sep = comma]{images/WGjunctionS21_GaussSep25.csv};
        \legend{AAA, VF, Sz.-Rat.,Szegö, SE}
      \end{semilogyaxis}
    \end{tikzpicture}
  \end{subfigure}
  \begin{subfigure}[b]{.46\textwidth}
    \centering
    \begin{tikzpicture}
      \begin{semilogyaxis}[grid, xlabel={Number of training points
          $n$}, title={$S_{41}$ - RMSE}, width=5cm, height=4cm, legend
        pos = outer north east, legend style = {font=\small}, very
        thick, xmin=5, xmax=15, no marks, , title style =
        {yshift=-.5em}]
        \addplot[TUDa-1b] table[x=N, y=RMSE, col sep = comma]{images/WGjunctionS41_AAA.csv};
        \addplot[TUDa-7b] table[x=N, y=RMSE, col sep = comma]{images/WGjunctionS41_VF.csv};
        \addplot[TUDa-9b] table[x=N, y=RMSE, col sep = comma]{images/WGjunctionS41_Adap.csv};
        \addplot[TUDa-4d, densely dashed] table[x=N, y=RMSE, col sep = comma]{images/WGjunctionS41_Szego.csv};
        \addplot[TUDa-10b, densely dotted] table[x=N, y=RMSE, col sep = comma]{images/WGjunctionS41_GaussSep25.csv};
      \end{semilogyaxis}
    \end{tikzpicture}
  \end{subfigure}
  \caption{%
    Top left: Waveguide junction model, taken from CST Microwave
    Studio \cite{CST_2019aa}. %
    Top right: Complex frequency response functions. %
    Bottom: Convergence studies w.r.t.\ the number of training points.}
  \label{fig:WGjunction}
\end{figure}

Next, we consider an electromagnetic model problem, which is a
demonstration example of CST Microwave Studio \cite{CST_2019aa},
solving the full set of Maxwell equations in the frequency domain. %
The model consists of a waveguide junction with 4 ports, which
contains a small metallic disk and is connected to an external cavity
resonator (see Figure~\ref{fig:WGjunction}). %
The structure is excited at the first port and simulated using the
finite element method in the frequency domain. %
In particular, we set the solver accuracy of the 3rd order solver to
$10^{-6}$ and use a curved mesh with standard settings. %
We employ an initial adaptive mesh refinement at \SI{9}{GHz}, where we
set the scattering parameter 
criterion threshold with 2 subsequent checks to
$10^{-4}$. %
As quantities of interest we consider the scattering parameters on a
frequency range of [\SI{7}{GHz},\,\SI{9}{GHz}] using equidistant
sample points, where we restrict ourself to $S_{21}$ and $S_{41}$ for
brevity, however, the results are qualitatively similar for all four
scattering parameters. %
It can be seen that the quantities of interest have a dominant pole at around
\SI{8}{GHz}. %
This causes the purely kernel-based interpolations to be inferior
compared to the rational approximations. %
However, the proposed combination of kernel-based interpolation and
rational approximations leads to satisfactory results, with an
accuracy comparable to that of~AAA and~VF.%

\begin{figure}
  \begin{subfigure}[b]{.49\textwidth}
    \centering
    \tdplotsetmaincoords{70}{115}	
    \scriptsize
    \begin{tikzpicture}[scale=1.8, tdplot_main_coords]
      \pgfmathsetmacro{\cubex}{2}
      \pgfmathsetmacro{\cubey}{1.5}
      \pgfmathsetmacro{\cubez}{1}

      \draw[black, thick, fill = TUDa-1b!40] (0,0,\cubez) -- (\cubex,0,\cubez) -- (\cubex,\cubey,\cubez) -- (0,\cubey,\cubez) -- cycle;	
      \draw[black, thick, fill = TUDa-1b!40] (0,\cubey,0) -- (0,\cubey,\cubez) -- (\cubex,\cubey,\cubez) -- (\cubex,\cubey,0) -- cycle;	
      \draw[black, thick, dotted] (0,0,0) -- (0,\cubey,0) -- (0,\cubey,\cubez) -- (0,0,\cubez) -- cycle;
      \draw[black, thick, fill = gray!50] (\cubex,0,0) -- (\cubex,\cubey,0) -- (\cubex,\cubey,\cubez) -- (\cubex,0,\cubez) -- cycle;

      \tikzset{>=latex}
      \draw[ ->,  very thick] (\cubex+.7,0,0) -- (\cubex,0,0);
      \node[below right] at (\cubex+.7, 0,0) {\small Point force $F_s$};
		
      \node[inner sep=2pt, rectangle, fill=white, draw=black, text centered] at (\cubex/2, \cubey/2, \cubez) {\small Air cavity $D_f$};
      \node[inner sep=2pt, rectangle, fill=white, draw=black, text centered] at (\cubex, \cubey/2, \cubez/2) {\small Plate $D_s$};
    \end{tikzpicture}
    \vspace{1em}
  \end{subfigure}
  \hfill
  \begin{subfigure}[b]{.49\textwidth}
    \centering
    \begin{tikzpicture}
      \begin{axis}[grid, xlabel={$\omega$ (rad$\mathrm{s}^{-1}$)},
        width=\textwidth, height=.7\textwidth, ylabel near ticks,
        legend style = {font=\small}, legend pos = north east, xmin =
        4500, xmax=5000, ymax=5.2, ymin=-2.8, xtick={4500, 4750,
          5000}]
        \addplot[black, very thick, no marks] table[x=x,y expr={\thisrow{y_ref_real}*1e8}, col sep = comma]{images/illustration_VibroAcoustics.csv};
        \addplot[TUDa-9b, very thick, no marks] table[x=x, y expr={\thisrow{y_ref_imag}*1e8}, col sep = comma]{images/illustration_VibroAcoustics.csv};
        \addplot[TUDa-1b, dashed, very thick, no marks] table[x=x, y expr={\thisrow{y_ref_abs}*1e8}, col sep = comma]{images/illustration_VibroAcoustics.csv};
        \legend{Real part, Imag. part, Magnitude}
      \end{axis}
    \end{tikzpicture}
  \end{subfigure}
  \begin{subfigure}[b]{1\textwidth}
    \centering
    \begin{tikzpicture}
      \begin{semilogyaxis}[grid, xlabel={Number of training points
          $n$}, ylabel={RMSE}, width=.6\textwidth,
        height=.33\textwidth, legend pos = outer north east, legend
        style = {font=\small}, ytick = {1e-1, 1e-2, 1e-3, 1e-4, 1e-5,
          1e-6}, no marks, very thick, xmin=15, every axis
        plot/.append style={very thick},xmax=50]
        \addplot[TUDa-1b] table[x=N, y=RMSE, col sep = comma]{images/VibroAcoustics_AAA.csv};
        \addplot[TUDa-7b] table[x=N, y=RMSE, col sep = comma]{images/VibroAcoustics_VF.csv};
        \addplot[TUDa-9b] table[x=N, y=RMSE, col sep = comma]{images/VibroAcoustics_Adap.csv};
        \addplot[TUDa-4d, densely dashed] table[x=N, y=RMSE, col sep = comma]{images/VibroAcoustics_Szego.csv};
        \addplot[TUDa-10b, densely dotted] table[x=N, y=RMSE, col sep = comma]{images/VibroAcoustics_GaussSep.csv};
        \legend{AAA, VF, Sz.-Rat., Szegö, SE}
      \end{semilogyaxis}
    \end{tikzpicture}
  \end{subfigure}
  \caption{%
    Top left: Vibro-acoustic benchmark problem, based on
    \cite{Roemer_2021aa}. %
    Top right: Complex frequency response function. %
    Bottom: Convergence study w.r.t.\ the number of training points.}
  \label{fig:Vibroacoustic}
\end{figure}
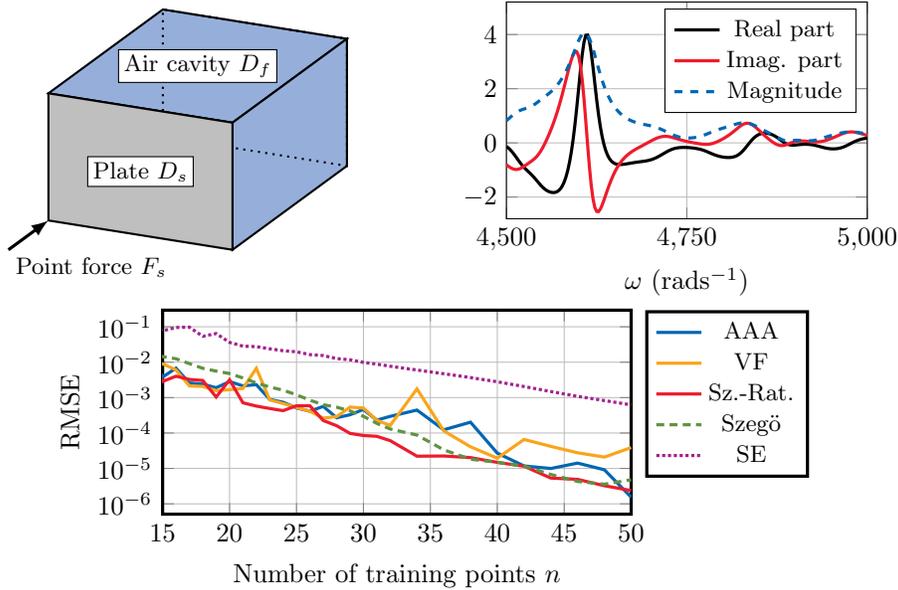

The final test case is a vibroacoustic finite element model, taken
from \cite{Roemer_2021aa} and depicted in
Figure~\ref{fig:Vibroacoustic}. %
A 2D Mindlin plate (vibrating structure $D_s$) is excited by a point
force and strongly coupled to a 3D acoustic domain (air
cavity~$D_f$). %
Then, the response at a point in the fluid is evaluated. %
See \cite{Roemer_2021aa} for more details on the model. %
We consider the frequency response on a frequency interval
$\omega \in [\SI{4500}{\per\second},\SI{5000}{\per\second}]$, shown in
Figure~\ref{fig:Vibroacoustic} (top, right). %
The convergence study, given in Figure~\ref{fig:Vibroacoustic} (bottom),
indicates that the proposed approach usually achieves an accuracy at
least comparable to that of~AAA and~VF with, at certain points, an
observable improvement by about an order of magnitude. %
It can also be seen that the rational mean function improves the
accuracy at the majority of points compared to the pure Szegö
kernel-based interpolation.%

\section{Conclusion}
\label{sec:conclusions}
We have presented a comprehensive framework for kernel-based
interpolation of complex-valued functions and frequency response
functions. %
In the complex-valued case, the pseudo-kernel is an additional
ingredient, which can be used to improve the interpolation accuracy. %
We have introduced the concept of complex/real reproducing kernel
Hibert spaces to reveal the role of the pseudo-kernel and to establish
results on minimum norm interpolation. %
Furthermore, we have proposed a hybrid method, which complements the
kernel-interpolant with a low-order rational function and a new model
selection criterion: this extension is crucial to account for dominant
poles in applications.

The capabilities of the rational-kernel method have been illustrated
with several examples, from circuits to frequency response functions
originating from PDE problems. %
In all examples the performance was at least comparable, in some cases
improved, compared to AAA and vector fitting on the same set of
training data.

The kernel method was further linked to complex-valued Gaussian
process regression, which can be used in future work to include noise, quantify uncertainty and carry out adaptive
  sampling. To this end, it will be important to correctly account for
  uncertainties in the estimation of the low-order rational mean
  function, i.e., uncertainties in the number of poles and the estimated poles and residues.
A generalization to the multivariate case, where, e.g., uncertain
parameters are considered as well, and comparisons against
multivariate AAA \cite{rodriguez2020p} or rational Polynomial Chaos
\cite{schneider2023sparse}, would also be of interest.

\appendix
\section{Proofs}
\label{sec:proofs-appendix}

\subsection{Proof of Theorem~\ref{thm:gammaAlpha}}
\label{proof:thmGammaAlpha}

We assume without loss of generality that~$\alpha = 0$ in this
proof---i.e., we consider the case of the Hardy space~$H^2(\Gamma_0)$
on the right
half-plane~$\Gamma_0 = \left\{ s \in \mathbb{C} \,\mid\, \Re[s] >0
\right\}$. %
The general case follows by translation.

The fact that~$H^2(\Gamma_0)$ is an RKHS is well known. %
Indeed, recall the one-sided Paley-Wiener theorem
(see, e.g., Chapter~8 of~\cite{hoffman:1962}): %
for all~$f \in H^2(\Gamma_0)$, there exists a
unique~$\widehat f \in L^2(\Rset_+)$ such that
\begin{equation}\label{equ:PW-repr}
  f(s) \;=\; \frac{1}{\sqrt{2\pi}} \int_0^{+\infty} \widehat f(t)\, e^{-st}\, \dt,
  \qquad \forall s \in \Gamma_0,
\end{equation}
and the mapping $f \mapsto \widehat f$ is a surjective isometry:
$\lVert f \rVert_{H^2(\Gamma_0)} = \lVert \widehat f
\rVert_{L^2(\Rset_+)}$. %
This proves that $H^2(\Gamma_0)$~is a Hilbert space, and a simple
application of the Cauchy-Schwarz inequality for
$s = x+\i y \in \Gamma_0$ yields:
\begin{equation*}
  \left| f(s) \right| \;\le\;
  \frac{1}{2\sqrt{\pi x}} \cdot \lVert \widehat f \rVert_{L^2(\Rset_+)},
\end{equation*}
which proves that the evaluation functionals are continuous
on~$H^2(\Gamma_0)$.

Let us now determine the kernel~$k$ of this RKHS. %
Let $s_0 \in \Gamma_0$ and set $h = k(\cdot, s_0)$. %
Then, for any~$f \in H^2(\Gamma_0)$, the reproduction property
combined with~\eqref{equ:PW-repr} yields:
\begin{equation*}
  \left< f,\, h \right>_{H^2(\Gamma_0)}
  \;=\; f(s_0)
  \;=\; \frac{1}{\sqrt{2\pi}} \int_0^{+\infty} \widehat f(t)\, e^{-s_0t}\, \dt
  \;=\; \left< \widehat f,\, \frac{1}{\sqrt{2\pi}}\, e^{-s_0^* (\cdot)} \right>_{L^2(\Rset_+)},
\end{equation*}
which implies that
$\widehat h = \frac{1}{\sqrt{2\pi}}\, e^{-s_0^* (\cdot)}$ since
$f \mapsto \widehat f$ is an isometric isomorphism. %
The expression of the kernel follows:
\begin{equation}\label{equ:kernel:Gamma0}
  k(s, s_0) \;=\; h(s)
  \;=\; \frac{1}{\sqrt{2\pi}} \int_0^{+\infty} \widehat{h}(t)\, e^{-st}\, \dt
  \;=\; \frac{1}{2\pi \left( s + s_0^* \right)}.
\end{equation}

It remains to show that $k$~is strictly positive definite. %
For any $m \ge 1$ and $s_1, \ldots, s_m \in \Gamma_0$, the kernel
matrix $K_m = \left( k(s_i, s_j) \right)_{1 \le i, j \le m}$ can be
seen as the conjugate Gram matrix of~$h_1, \ldots, h_m$
in~$L^2(\Rset_+)$, where
$h_j(t) = \frac{1}{\sqrt{2\pi}}\, e^{-s_j^* t}$, $t \ge 0$. %
Assume that $s_1$, \ldots, $s_m$ are distinct. %
Then it is well known that the complex
exponentials~$e^{-s_1^* (\cdot)}$, \ldots, $e^{-s_m^* (\cdot)}$ are
linearly independent entire functions on~$\Cset$. %
It follows, using the identity theorem, that $h_1$, \ldots, $h_m$ are
linearly independent as well. %
The kernel matrix $K_m$~is thus invertible and, consequently, positive
definite. %
Therefore $k$ is strictly positive definite.

\begin{remark} 
  The expression of the reproducing kernel is also derived in
  \cite[Theorem 2.12]{Bonyo_2020aa} (for the upper half-plane instead
  of~$\Gamma_0$) using a different approach involving the kernel of
  the Hardy space of the unit disk. %
  Note, however, that the factor~$2\pi$ in the denominator
  of~\eqref{equ:kernel:Gamma0} is missing
  in~\cite[Equation~(2.9)]{Bonyo_2020aa}; %
  the discrepancy comes from a missing factor~$\frac{1}{2\pi}$ in the
  definition of the norm on~$H^p(\Dset)$ on page~14.
\end{remark}

\subsection{Proof of Proposition~\ref{prop:counterexample-dim2}}

Take $H = \left\{ \alpha f_0,\, \alpha \in \Cset \right\}$, where
$f_0:\Xset \to \Cset$ is some fixed function, and define a real inner
product over~$H$ by
$\left< \alpha f_0,\, \beta f_0 \right> := \Re\alpha \cdot \Re\beta +
4\, \Im\alpha \cdot \Im \beta$.  %
Assuming that $f_0 \not\equiv 0$, the resulting space is complex/real
RKHS of dimension two, spanned by~$\left\{ f_0,\, \i f_0 \right\}$. %
($H$ is also a complex vector space of dimension~$1$.)

It not possible to embed~$H$ as a subspace of a complex Hilbert
space~$H_\Cset$ with inner product~$\left< \cdot, \cdot \right>_\Cset$
such that $\left<f, g \right> = \Re \left<f, g \right>_\Cset$ for all
$f, g \in H$. %
To see it, note for instance that $\lVert f_0 \rVert = 1$ while
$\lVert \i f_0 \rVert = 2$.

\subsection{Proof of Proposition~\ref{prop:repr-eval-func}}
\label{proof:propReprEvalFunc}

Let $f \in H$, $s_0 \in \Sset$ and~$a_0 \in \{ \re, \im \}$. Then
\begin{align}
  G_{a_0} \left( f(s_0) \right)
  &\;=\; \left( \Acal f \right)(s_0, a_0)
  \;=\; \left< \Acal f,\, \tilde k\left( \cdot,\, (s_0, a_0)\right) \right>_{\tilde H}\\
  &\;=\; \left< f,\, \Acal^{-1} \left( \tilde k\left( \cdot,\, (s_0, a_0)\right) \right) \right>_H.
    \label{equ:proof:repr-eval-func:2}
\end{align}
Taking $a_0 = \re$, we have thus proved that
$\Re \circ \delta_{s_0} = \left<\, \bm{\cdot}\,,\, \varphi_\re \left(
    \cdot, s_0 \right) \,\right>_H$, where
\begin{equation}
  \varphi_\re \left( \cdot, s_0 \right) = \Acal^{-1} \left( %
    \tilde k\left( \cdot,\, (s_0, \re)\right) %
  \right) \in H
\end{equation}
can be computed as follows:
\begin{align}
  \Re \left[ \varphi_\re \left( s, s_0 \right) \right]
  & \;=\; \left( \Acal\left[ \varphi_\re \left( \cdot, s_0 \right) \right] \right)(s, \re)
    \;=\; \tilde k\left( (s, \re),\, (s_0, \re)\right)
    \;=\; k_{\re\re}(s, s_0),\\
  \Im \left[ \varphi_\re \left( s, s_0 \right) \right]
  & \;=\; \left( \Acal\left[ \varphi_\re \left( \cdot, s_0 \right) \right] \right)(s, \im)
    \;=\; \tilde k\left( (s, \im),\, (s_0, \re)\right)
    \;=\; k_{\im\re}(s, s_0).
\end{align}
The expression of~$\varphi_\im \left( \cdot, s_0 \right)$ is derived similarly
by taking $a_0 = \im$ in~\eqref{equ:proof:repr-eval-func:2}.

\subsection{Proof of Proposition~\ref{prop:dense-subspace}}

In a real or complex RKHS, it is well known that the partial kernel
functions~$k(\cdot,\, s_0)$, $s \in \Sset$, span a dense subset of the
Hilbert space. %
Moreover, recall that the bijection~$\Acal$ defined in
Section~\ref{sec:complex-real-RKHS} is an isometric isomorphism
between~$H$ and a real RKHS $\tilde H$
on~$\tilde\Sset = \Sset \times \{\re,\im\}$, whose kernel~$\tilde k$
can be recovered from~$k$ and~$c$ by inverting
\eqref{prop:repr-eval-func}--\eqref{def:complex-kernels}. %
The claim then follows from the observation that any function
on~$\tilde\Sset$ of the form
\begin{equation*}
  \tilde g \;=\;
  \sum_{i=1}^n \alpha_i\, \tilde k\left( \cdot, (s_i, \re) \right)
  + \sum_{i=1}^n \beta_i\, \tilde k\left( \cdot, (s_i, \im) \right),
\end{equation*}
where $\alpha_1, \beta_1, \ldots, \alpha_n, \beta_n \in \Rset$,
corresponds to the image by~$\Acal$ of
\begin{align*}
  g & \;=\;   \sum_{i=1}^n \alpha_i\, \Acal^{-1}\left( \tilde k\left( \cdot, (s_i, \re) \right) \right)
      + \sum_{i=1}^n \beta_i\, \Acal^{-1} \left( \tilde k\left( \cdot, (s_i, \im) \right) \right)\\
    & \;=\;   \sum_{i=1}^n \alpha_i\, \varphi_\re\left( \cdot, s_i \right)
      + \sum_{i=1}^n \beta_i\, \varphi_\im\left( \cdot, s_i \right)\\
    & \;=\;   \sum_{i=1}^n \gamma_i\, k\left( \cdot, s_i \right)
      + \sum_{i=1}^n \gamma_i^*\, c\left( \cdot, s_i \right),
      \qquad \text{with } \gamma_i = \frac{1}{2} \left( \alpha_i + \i \beta_i \right).
\end{align*}

\subsection{Proof of Theorem~\ref{thm:CR-RKHS-characterization}}

Assume first that~$k$ and~$c$ are the complex kernel and pseudo-kernel
associated to a given complex/real RKHS~$H$. %
Let $\tilde\xi$ denote a zero-mean (e.g., Gaussian) real-valued random
process indexed by~$\Sset$ with covariance function equal to the
kernel~$\tilde k$ of the real RKHS~$\tilde H = \Acal H$, and set
$\xi = \tilde\xi(\cdot,\re) + \i\, \tilde\xi(\cdot,\im)$. %
Then $\xi$ is a complex-valued random process on~$\Sset$, with
covariance function~$k$ and pseudo-covariance function~$c$; %
indeed, for all $s, s_0 \in \Sset$,
\begin{align*}
  \Esp\left( \xi(s)\, \xi(s_0)^* \right)
  & \;=\; \left(
    \tilde k\left( (s,\re),\, (s_0, \re) \right)
    + \tilde k\left( (s,\im),\, (s_0, \im) \right)
    \right)\\
  & \quad + \i\, \left(
    \tilde k\left( (s,\im),\, (s_0, \re) \right)
    - \tilde k\left( (s,\re),\, (s_0, \im) \right)
    \right) \;=\; k(s, s_0),
\end{align*}
and similarly~$\Esp\left( \xi(s)\, \xi(s_0) \right) = c(s, s_0)$. %
It follows readily that $k$~is Hermitian and positive definite,
and that $c$ is symmetric, which proves i) and~ii).

Pick~$s_1, \ldots s_n \in \Sset$, and set
$K_n = \left( k(s_i,s_j) \right)_{1 \le i,j \le n}$ and
$C_n = \left( c(s_i,s_j) \right)_{1 \le i,j \le n}$. %
Then~$K_n$ and~$C_n$ are respectively the covariance and
pseudo-covariance matrix of the random
vector~$Z = \left( \xi(s_1),\, \ldots,\, \xi(s_n) \right)^\tra$, and
thus iv) is precisely the ``only if'' part the following result, due
to~\cite{picinbono:1996}.

\begin{proposition} \label{prop:Picinbono}
  Let $n \in \Nset^*$.  Let $K$ be a complex, Hermitian, positive
  definite matrix of order~$n$, and let $C$ be a complex, symmetric
  matrix of the same size.  Then there exists a complex random
  vector~$Z$ with covariance matrix~$K$ and pseudo-covariance
  matrix~$C$ if, and only if, $K^* - C^\her K^{-1} C$ is positive
  semi-definite.
\end{proposition}

It remains to prove~iii): let $u \in \ker K_n$. %
Then $u^\her K_n u = \Esp\left( \left| u^\her Z \right|^2 \right) = 0$,
therefore $u^\her Z = 0$ almost surely, and as a consequence:
\begin{equation*}
  C_n^* u
  \;=\; \Esp\left( Z Z^\tra \right)^*\, u
  \;=\; \Esp\left( Z^* Z^\her u \right)
  \;=\; \Esp\left( Z^* (u^\her Z)^\her \right)
  \;=\; 0.
\end{equation*}
This completes the proof of~i)--iv).

Conversely, assume now that~$k$ and~$c$ are two functions
from~$\Sset \times \Sset$ to~$\Cset$, such that i)--iv) hold. %
Then it is easy to see that there is a unique
function~$\tilde k: \Sset \times \{ \re, \im \} \to \Rset$ such that
\eqref{eq:complex-kernel}--\eqref{eq:pseudo-kernel} hold, given by
\begin{align*}
  k_{\re\re}(s, s_0) & \;=\; \frac{1}{2}\,  \Re \left( k(s,s_0) + c(s,s_0) \right)\\
  k_{\im\im}(s, s_0) & \;=\; \frac{1}{2}\,  \Re \left( k(s,s_0) - c(s,s_0) \right)\\
  k_{\im\re}(s, s_0) & \;=\; \frac{1}{2}\,  \Im \left( k(s,s_0) + c(s,s_0) \right)
                      \;=\; k_{\re\im}(s_0, s).
\end{align*}
It remains to prove that~$\tilde k$ is positive definite. %
It is easy to see that this is true if, and only if, the
matrices~$K_n$ and~$C_n$ defined above are the covariance and
pseudo-covariance matrices of a complex random vector~$Z$, for any
choice of the points~$s_1, \ldots, s_n \in \Sset$. %
Pick such a set of points, and let $r$ denote the rank of~$K_n$. %
Assume without loss of generality that
\begin{equation}
  \label{eq:Kn-block}
  K_n \;=\;
  \begin{pmatrix}
    K_{11} & K_{12}\\
    K_{12}^\her & K_{22}
  \end{pmatrix},
\end{equation}
with $K_{11}$ a positive definite $r \times r$ matrix. %
Then $K_{22} = K_{12}^\her K_{11}^{-1} K_{12}$ and
\begin{equation}
  \label{eq:Kn-block-bis}
  K_n \;=\; M\,
  \begin{pmatrix}
    K_{11} & 0\\
    0 & 0
  \end{pmatrix}\,
  M^\her,
  \qquad \text{where }
  M \;=\;
  \begin{pmatrix}
    \mathrm{I}_r & 0\\
    K_{12}^\her K_{11}^{-1} & \mathrm{I}_{n-r}
  \end{pmatrix}
\end{equation}
Denote by~$C_{11}$ the upper-left $r \times r$ block in~$C_n$. %
Then it follows from iv) that
$K_{11}^* - C_{11}^\her K_{11}^{-1} C_{11}$ is positive semi-definite,
and thus by Proposition~\ref{prop:Picinbono} there exists a complex
random vector~$Z_1$ of size~$r$ with covariance matrix~$K_{11}$ and
pseudo-covariance matrix~$C_{11}$. %
It is then clear from~\eqref{eq:Kn-block-bis} that $K_n$ is the
covariance matrix of
\begin{equation*}
  Z = M\,
  \begin{pmatrix}
    Z_1\\ 0
  \end{pmatrix}.
\end{equation*}
To complete the proof, it remains to observe that~$C_n$ is the
pseudo-covariance matrix of~$Z$:
\begin{equation}
  \label{eq:Cn-block}
  C_n \;=\; M\,
  \begin{pmatrix}
    C_{11} & 0\\
    0 & 0
  \end{pmatrix}\, M^\tra \;=\; \Esp\left( Z Z^\tra \right),
\end{equation}
which follows from the facts that~$C_n$ is symmetric and that
$\ker K_n \subset \ker C_n^*$, respectively by~ii) and~iii).

\subsection{Proof of Theorem~\ref{thm:interp:cr}} %
Using the bijection~$\Acal$ defined in
Section~\ref{sec:complex-real-RKHS}, the interpolation problem
on~$\Sset$ with complex-valued data~$\left( s_1, y_1 \right)$, \ldots,
$\left( s_n, y_n \right)$ can be reformulated as an interpolation
problem on~$\tilde\Sset = \Sset \times \{\re,\im\}$ with real-valued
data~$\left( (s_1, \re), \Re(y_1) \right)$,
$\left( (s_1, \im), \Im(y_1) \right)$, \ldots,
$\left( (s_n, \re), \Re(y_n) \right)$,
$\left( (s_n, \im), \Im(y_n) \right)$. %
The claim then follows from Theorem~\ref{thm:complexRKHSinterpolation}
using, as in the proof of Proposition~\ref{prop:dense-subspace}, the
fact that $\Acal$ is an isometric isomorphism between~$H$ and the real
RKHS $\tilde H = \Acal(H)$.

\subsection{Proof of Theorem~\ref{thm:hermitian}} %
$i) \Rightarrow ii)$. %
Let $H$ denote a complex/real RKHS on~$\Sset$ with complex kernel~$k$,
such that \eqref{equ:symmetry} holds. %
Let $c$ denote the pseudo-covariance of~$H$. %
Let $s_0 \in \Sset$. %
It follows from Proposition~\ref{prop:dense-subspace} that
\begin{equation*}
  f_\gamma \;=\; \gamma\, k(\cdot, s_0) \,+\, \gamma^*\, c(\cdot, s_0)
\end{equation*}
is in~$H$ for all~$\gamma \in \Cset$. %
Using~\eqref{equ:symmetry}, we see then that
\begin{align*}
  f_\gamma(s^*) & \;=\; \gamma\, k(s^*, s_0) \,+\, \gamma^*\, c(s^*, s_0)\\
         & \;=\; \gamma\, c(s, s_0)^* \,+\, \gamma^*\, k(s, s_0)^* \;=\; f_\gamma(s)^*
\end{align*}
holds for all~$\gamma \in \Cset$. %
This yields in particular that
$c(s,s_0) = k(s^*, s_0)^* = k(s_0, s^*)$, and the claim follows from
the symmetry of~$c$:
\begin{equation*}
  c(s, s_0) = c(s_0, s) = k(s, s_0^*).
\end{equation*}
Note that we have actually proved a little more than~$ii)$: if $i)$
holds, then $ii)$ holds for the \emph{same} complex/real RKHS~$H$. %
Since we will now prove that $ii) \Rightarrow iii) \Rightarrow i)$, it
follows that the complex/real RKHS with complex kernel~$k$ and
pseudo-kernel~$c$ defined by~\eqref{equ:pseudo-kern-symm}, if it
exists, is the only complex/real RKHS with complex kernel~$k$ such
that~\eqref{equ:symmetry} holds.

$ii) \Rightarrow iii)$. %
Let $H$ denote a complex/real RKHS on~$\Sset$ with complex kernel~$k$.
Assume that the pseudo-kernel $c$
satisfies~\eqref{equ:pseudo-kern-symm}. %
Then, for all $s, s_0 \in \Sset$,
\begin{equation*}
  k(s, s_0^*) \;=\; c(s, s_0) \;=\; c(s_0, s) \;=\; k(s_0, s^*).
\end{equation*}

\newcommand \ipC  {\left< \cdot,\, \cdot \right>_{\Cset}}
\newcommand \ipR  {\left< \cdot,\, \cdot \right>_{\Rset}}
\newcommand \truc {\diamond}

$iii) \Rightarrow i)$. %
Let $k$ denote a Hermitian positive definite kernel on~$\Sset$ such
that
\begin{equation}\label{equ:assumpt-iii}
  \forall s,s_0 \in \Sset, \quad k(s, s_0^*) = k(s_0, s^*).
\end{equation}
Let $(H_\Cset, \ipC)$ denote the complex RKHS with kernel~$k$ and let
$\ipR = \Re \ipC$. %
Then, as observed in Remark~\ref{rem:complex-RKHS-subspaces},
$(H_\Cset, \ipR)$ is a complex/real RKHS. %
The associated real and imaginary evaluation kernels, which we
denote by~$\varphi_\re^\truc$ and~$\varphi_\im^\truc$ respectively,
are easily seen to be given by~$\varphi_\re^\truc = k$
and~$\varphi_\im^\truc = \i\, k$, %
and the complex kernel and pseudo-kernel follow:
\begin{equation*}
  k^\truc = \varphi_\re^\truc - \i \varphi_\im^\truc = 2k
  \quad \text{and} \quad
  c^\truc = \varphi_\re^\truc + \i \varphi_\im^\truc = 0.
\end{equation*}

\bigbreak Now let~$H$ denote the subset of all the
functions~$f \in H_\Cset$ that satisfy~\eqref{equ:symmetry}: $H$ is
clearly a real subspace of~$H_\Cset$, and thus $\left( H, \ipR \right)$
is a complex/real RKHS as well. %
Moreover, for any $f \in H$,
\begin{align*}
  \Re\, f(s)
  & \;=\; \Re\, \left\{
    \frac{1}{2}\, \left( f(s) + f(s^*)^* \right) \right\}\\
  & \;=\; \frac{1}{2}\, \Bigl\{
    \left< f,\, \varphi_\re^\truc(\cdot,s) \right>_\Rset
    + \left< f,\, \varphi_\re^\truc(\cdot,s^*) \right>_\Rset
    \Bigr\}\\
  & \;=\; \left< f,\,
    \frac{1}{2}\, \left(
    \varphi_\re^\truc(\cdot,s) + \varphi_\re^\truc(\cdot,s^*) \right)
    \right>_\Rset.
\end{align*}
As a consequence of~\eqref{equ:assumpt-iii}, the function
$\frac{1}{2}\, \left( \varphi_\re^\truc(\cdot,s) +
  \varphi_\re^\truc(\cdot,s^*) \right)$ in this inner product satisfies, for all $s_0 \in \Sset$,
\begin{equation*}
		\frac{1}{2}\, \left( \varphi_\re^\truc(s_0,s)^* +
  \varphi_\re^\truc(s_0,s^*)^* \right)
  	\;=\; \frac{1}{2}\, \left( \varphi_\re^\truc(s,s_0) +
  \varphi_\re^\truc(s^*,s_0) \right).
\end{equation*}
Hence, it is an element of~$H$, which proves that the real evaluation
functional~$\varphi_\re$ of~$\left( H, \ipR \right)$ is given by
\begin{equation*}
  \varphi_\re(s, s_0)
  \;=\; \frac{1}{2}\, \left(
    \varphi_\re^\truc(s,s_0) + \varphi_\re^\truc(s,s_0^*) \right)
  \;=\; \frac{1}{2}\, \left(  k(s,s_0) + k(s,s_0^*) \right).
\end{equation*}
Similarly for the imaginary evaluation functional~$\varphi_\im$:
\begin{equation*}
  \varphi_\im(s, s_0)
  \;=\; \frac{1}{2}\, \left(
    \varphi_\im^\truc(s,s_0) - \varphi_\im^\truc(s,s_0^*) \right)
  \;=\; \frac{\i}{2}\, \left(  k(s,s_0) - k(s,s_0^*) \right).
\end{equation*}
Therefore $\varphi_\re - \i \varphi_\im = k$ is the complex kernel
of~$\left( H, \ipR \right)$, which proves~$i)$.

To prove the remaining assertions, assume that~$i$--$iii)$ hold. %
Let $G$ denote the closed linear span
of~$\left\{ k(\cdot,s_0);\; s_0 \in \Sset \right\}$ over~$\Rset$. %
Then we have $G + \i G = H_\Cset$, and it follows
from~\eqref{equ:assumpt-iii} that~$G \subset H$. %
Observing that
\begin{equation*}
  \i H = \left\{
    f \in H_\Cset \mid \forall s \in \Sset,\, f(s^*) = - f(s)^*
  \right\},
\end{equation*}
we conclude that $H \cap \i H = \{ 0 \}$, therefore $G = H$ and
$H \oplus \i H = H_\Cset$. %

\subsection{Proof of Theorem~\ref{thm:existence-uniqueness-hermit}}
Observe first that, without loss of generality, we can add $m$~extra
data points~$(s_i, y_i)$, for some $m \le n$, in such way that 1) the
points~$s_i \in \Sset$ ($1 \le i \le n+m$) are still distinct, and 2)
for each~$i$ we have $s_j = s_i^*$ and~$y_j = y_i^*$ for some~$j$.

\paragraph{Existence} %
Since $k$ is strictly positive definite, we can find~$\alpha_1$,
\ldots, $\alpha_{n+m} \in \Cset$ such that
$h = \sum_{i=1}^{n+m} \alpha_i k(\cdot, s_i)$ interpolates the
extended data~$(s_1, y_1)$, \ldots, $(s_{n+m}, y_{n+m})$. %
This function $h$ belongs to~$H_\Cset$ but not in general to~$H$. %
Set $g(s) = \frac{1}{2} \left( h(s) + h(s^*)^* \right)$. %
Then $g$ clearly satisfies the symmetry condition ($g(s^*) = g(s)^*$
for all~$s \in \Sset$) and still interpolates the extended
data~$(s_1, y_1)$, \ldots, $(s_{n+m}, y_{n+m})$. %
Moreover, using iii) from Theorem~\ref{thm:hermitian}, we obtain that
\begin{equation*}
  g(s) = \frac{1}{2} \sum_{i=1}^{n+m}
  \left( \alpha_i k(s, s_i) + \alpha_i^* k(s, s_i^*) \right),
\end{equation*}
which shows that~$g \in H_\Cset$, and thus $g \in H$. %
Besides, we easily see using~\eqref{equ:pseudo-kern-symm} that: if
$s_i = s_i^*$ then
\begin{equation}
  \frac{1}{2} \Bigl\{
    \alpha_i k(s, s_i) + \alpha_i^* k(s, s_i^*)
  \Bigr\}
  =
  \gamma_i k(s, s_i) + \gamma_i^* c(s, s_i)
\end{equation}
with $\gamma_i = \frac{1}{2} \alpha_i$, and if $s_j = s_i^*$ with
$i \neq j$ then
\begin{equation} \label{equ:trotro}
  \begin{gathered}
    \frac{1}{2} \Bigl\{
    \bigl( \alpha_i k(s, s_i) + \alpha_i^* k(s, s_i^*) \bigr)
    +
    \bigl( \alpha_j k(s, s_j) + \alpha_j^* k(s, s_j^*) \bigr)
    \Bigr\}\\
    =
    \bigl(
      \gamma_i k(s, s_i) + \gamma_i^* c(s, s_i)
    \bigr) + \bigl(
      \gamma_j k(s, s_j) + \gamma_j^* c(s, s_j)
    \bigr)
  \end{gathered}  
\end{equation}
with $\gamma_i = \frac{1}{2} \left( \alpha_i + \alpha_j^* \right)$
and~$\gamma_j = 0$. %
It follows that $g$ can be rewritten under the
form~\eqref{equ:interpolant:cr}, using the fact that~$\gamma_j = 0$
in~\eqref{equ:trotro} to get rid of the $m$ extra terms. %
Thus $\gamma = \left( \gamma_1, \ldots, \gamma_n \right)^\tra$
solves~\eqref{equ:interp-system:cr}, %
which proves the ``existence'' part of the theorem.

\paragraph{Uniqueness} %
Let $g \in H$ denote a function of the
form~\eqref{equ:interpolant:cr}, where the coefficients~$\gamma_i$ are
such that \eqref{equ:interp-system:cr}~holds. %
Using the property that $c(s, s_i) = k(s, s_i^*)$, any such function
can be rewritten as $g = \sum_{i=1}^{n+m} \alpha_i k(\cdot, s_i)$. %
Moreover, since the $s_i$'s are $n + m$ distinct points in~$\Sset$ and
$k$~is strictly positive definite, the
coefficients~$\alpha_i \in \Cset$ are uniquely determined by the
interpolation conditions: $g(s_i) = y_i$, $1 \le i \le n+m$. %
The first $n$ conditions come directly
from~\eqref{equ:interp-system:cr}, and the $m$ additional conditions
must hold as well by symmetry, since $g \in H$. %

For each~$i$ such that $s_i = s_i^*$, it is easily seen that
$\alpha_i = \gamma_i + \gamma_i^*$ is real, and thus the value
of~$\gamma_i$ is uniquely determined by~$\alpha_i$ and the additional
condition that~$\gamma_i = \gamma_i^*$. %
Similarly, if $s_i = s_j^*$ for some $i, j \le n$, $i \neq j$, then
$\alpha_i = \gamma_i + \gamma_j^*$,
$\alpha_j = \gamma_i^* + \gamma_j$, and therefore $\gamma_i$,
$\gamma_j$ are uniquely determined by $\alpha_i$, $\alpha_j$ and the
condition~$\gamma_i = \gamma_j^*$. %
Finally, if $s_i = s_j^*$ for some $i \le n$ and $j > n$, then
$\alpha_i = \gamma_i$. %
We have thus proved that there is a unique
$\gamma = \left( \gamma_1, \ldots, \gamma_n \right)^\tra$, with the
property that $\gamma_i = \gamma_j^*$ when $s_i = s_j^*$, such that
\eqref{equ:interp-system:cr}~holds.

\bigskip

\section{Details on parameter optimization}
\label{sec:details-param-optim}

\subsection{Penalized log-likelihood criterion}
\label{sec:penalized-loglik}

We select the parameters~$\alpha$ and~$\sigma^2$ of the the Szegö
kernel~$\sigma^2 k_\alpha$ (cf.~Section~3 in the article), together
with the dominant poles $\mathbf p = \left( p_1, \ldots, p_K \right)$
in the case of the hybrid method, by maximizing a penalized
log-likelihood criterion:
\begin{equation}\label{equ:penalized-loglik}
  J\left( \alpha, \sigma^2, \mathbf p \right)
  \;=\;  \max_{\mathbf r \in \Cset^K}
  \ln\left( p\left(\mathbf y|\alpha, \sigma^2, \mathbf p, \mathbf r\right) \right)
    + \ln(\rho(\alpha)),
\end{equation}
where the first term is the log-likelihood of the model, maximized
(profiled) analytically with respect to the residues~$\mathbf r$ of
the rational mean function model~$m$, and the second term is a penalty
term, designed to pull~$\alpha$ away from~$0$.

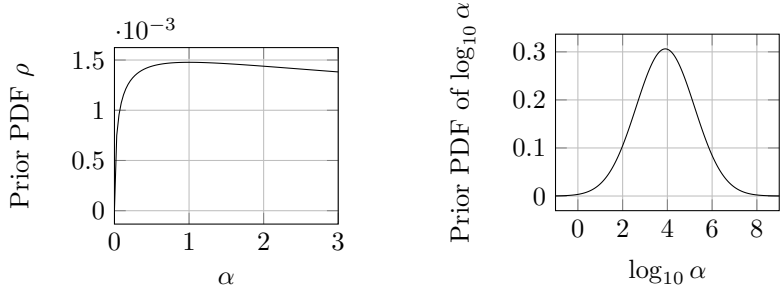
\begin{figure}
  \centering
  \begin{tikzpicture}
    \begin{axis}[width=.35\textwidth, grid, xlabel = {$\alpha$}, ylabel = {Prior PDF $\rho$}, xmin=0, xmax=3]
      \addplot[no marks] table[x=beta, y=pdf, col sep = comma]{images/PriorPDF.csv};
    \end{axis}
  \end{tikzpicture}\hspace{3em}
  \begin{tikzpicture}
    \begin{axis}[width=.35\textwidth, grid, xlabel = {$\log_{10}\alpha$}, ylabel = {Prior PDF of $\log_{10}\alpha$}, xmin=-1, xmax=9]
      \addplot[no marks] table[x=logalpha, y=pdf, col sep = comma]{images/PriorPDFlog.csv};
    \end{axis}
  \end{tikzpicture}
  \caption{%
    Log-normal prior on hyper-parameter $\alpha$ for $|\Omega|=1$.
    Left: log-normal prior density of~$\alpha$.  Note that the mode of
    the density is indeed at~$\alpha = \left| \Omega \right| = 1$. %
    Right: prior density of~$\log_{10}(\alpha)$. %
    This is a Gaussian density with
    mean~$\mu_\alpha / \ln(10) \approx 3.91$ and standard
    deviation~$\sigma_\alpha / \ln(10) \approx 1.30$.%
  }
  \label{fig:Prior}
\end{figure}

More precisely, we take for~$\rho(\alpha)$ the probability density
function (pdf) of a log-normal random variable with
parameters~$\mu_{\alpha}$ (to be specified) and~$\sigma_\alpha = 3$; %
in other words, we use a ``vague'' prior distribution on~$\alpha$,
such that~$\log(\alpha)$ is Gaussian with mean~$\mu_\alpha$ and
variance~$\sigma_\alpha^2$. %
The parameter~$\mu_\alpha$ is chosen in such a way that the log-normal
density for~$\alpha$ has its mode
at~$\left| \Omega \right| = \omega_{\mathrm{max}}
-\omega_{\mathrm{min}}$, or equivalently that the prior density
for~$\alpha / \left| \Omega \right|$ has its mode at~$1$. %
Using that the mode of the lognormal density is
at~$e^{\mu_\alpha - \sigma_\alpha^2}$, we deduce that
$\mu_\alpha = \sigma_\alpha^2 + \ln \left| \Omega \right|$. %
The resulting pdf
\begin{equation}
  \rho(\alpha) \;=\; \frac 1{\alpha\sigma_\alpha\sqrt{2\pi}}
  \exp\bigl(\frac {-(\ln(\alpha) -\mu_\alpha)^2}{2\sigma_\alpha^2}\bigr)
\end{equation}
is shown in Figure~\ref{fig:Prior} for~$\left| \Omega \right| = 1$. %
It can be seen that the chosen parameters allow for the choice
of~$\alpha$ within a range of several orders of magnitude.

The penalized log-likelihood criterion~\eqref{equ:penalized-loglik} is
maximized numerically using bound-constrained gradient-based
optimization---more precisely, interior point algorithm available from
Matlab's \texttt{fmincon} function---with a multistart procedure. %
Details about the bounds for the search domain and the initial points
for the local search are provided in
Sections~\ref{sec:bounds-search}--\ref{sec:starting-points}.

\begin{remark}
  This parameter selection procedure can be considered as
  \emph{maximum a posteriori} estimate in the Bayesian sense. %
  Indeed, the penalized log-likelihood
  criterion~\eqref{equ:penalized-loglik} can be seen as the
  log-posterior density, up to a constant, assuming a lognormal prior
  for~$\alpha$ and an improper uniform prior for all the other
  parameters.
\end{remark}

\begin{remark}%
  Even when the complex kernel~$k$ is strictly positive definite, the
  distribution of data under the GP model does not always admit a
  probability density function with respect to Lebesgue's measure
  on~$\Rset^{2n}$ (cf.~related discussion regarding the strict
  positive definiteness of~$\tilde k$ in Section~2.2 of the
  article). %
  When this happens, a suitable reference measure has to be used in
  order to define the likelihood function. %
  For instance, when the pseudo-kernel $c(s, s_0) = k(s, s_0^*)$ is
  used to enforce the symmetry condition, the value at~$\omega = 0$
  must be real, which yields a degenerate distribution if the response
  is evaluated at~$\omega = 0$: %
  the solution is simply to remove the imaginary part of the
  response at this point from the vector of observed variables. %
  See Section~2 of~\cite{lataire2016transfer} for related
  considerations.
\end{remark}

\subsection{Bounds for the search domain}
\label{sec:bounds-search}

We optimize with respect to the transformed kernel parameters
\begin{align*}
  \theta_1 & \;=\; \ln\left( \frac{\sigma^2}{2\pi} \right),\\
  \theta_2 & \;=\; \alpha,
\end{align*}
within the optimization bounds
\begin{align*}
  -15 & \;\le\; \theta_1 \;\le\; 15,\\
  10^{-6}|\Omega| & \;\le\; \theta_2 \;\le\; |\Omega|,
\end{align*}
where
$\left| \Omega \right| = \omega_{\max} - \omega_{\min} = \max \left\{
  \omega_i,\, 1 \le i \le n \right) - \min \left\{ \omega_i,\, 1 \le i
  \le n \right)$.

For the hybrid model, the poles are optimized simultaneously with the
kernel hyper-parameters, within the bounds
\begin{align*}
  -|\Omega| & \;\le\; \Re\left( p_i \right) \;\le\; -10^{-6}|\Omega|, \\
  \max\left\{10^{-6}|\Omega|,~\omega_{\min}-\frac{|\Omega|}3\right\}
            & \;\le\; \Im\left(p_i\right) \;\le\; \omega_{\max}+\frac{|\Omega|} 3.
\end{align*}
The bounds are enlarged, if needed, in such a way that all the poles
in the starting point of the optimization are contained within them.

\subsection{Starting point(s)}
\label{sec:starting-points}

We use a multistart procedure to optimize the kernel
parameters~$\alpha$ and~$\sigma^2$. %
More precisely, we start $N_\mathrm{ms} = 20$ local optimizations,
with the initial value of~$\alpha$ uniformly distributed
between~$10^{-6}|\Omega|$ and~$|\Omega|$. %
For a given value of~$\alpha$, and a given choice of poles in the case
of the hybrid algorithm, the GLS (generalized least squares) estimate
is used as a starting point for~$\sigma^2$.

For the poles in the hybrid algorithm, we start from equidistant
poles~$p_1, \ldots, p_K$ close to the frequency axis:
\begin{equation*}
  p_k \;=\; - \delta_{\Re}\, \left| \Omega \right|
  \,+\, \i \left( \omega_{\min} + \left( k -\frac{1}{2} \right)\, \delta_\Im \right),
  \qquad 1 \le k \le K,
\end{equation*}
where $\delta_{\Re} = 10^{-3}$ (weak attenuation) and
$\delta_\Im = \left| \Omega \right| / K_{\max}$. %
The kernel parameters are initialized a\textcolor{blue}{s} described previously (with the
GLS estimate for~$\sigma^2$ and a multi-start procedure
for~$\alpha$).

\section*{Acknowledgments}%
We thank Christopher Blech, Harikrishnan Sreekumar and Sabine Langer
for suggesting acoustic benchmarks and for providing the
implementation of the finite element solver used to compute the
vibroacoustics data set.

\phantomsection
\bibliographystyle{siamplain}
\bibliography{references}

\begin{thebibliography}{1}

\bibitem{bozzini:2013:generalized}
{\sc M.~Bozzini, M.~Rossini, and R.~Schaback}, {\em Generalized
  {W}hittle--{M}at{\'e}rn and polyharmonic kernels}, Advances in Computational
  Mathematics, 39 (2013), pp.~129--141.

\bibitem{hu:1998:collocation}
{\sc X.-G. Hu, T.-S. Ho, and H.~Rabitz}, {\em The collocation method based on a
  generalized inverse multiquadric basis for bound-state problems}, Computer
  physics communications, 113 (1998), pp.~168--179.

\bibitem{rasmussen:2006:gpml}
{\sc C.~E. Rasmussen and C.~K.~I. Williams}, {\em {Gaussian Processes for
  Machine Learning}}, {MIT} Press, 2006.

\bibitem{sollich:2004:using}
{\sc P.~Sollich and C.~Williams}, {\em Using the equivalent kernel to
  understand {G}aussian process regression}, Advances in Neural Information
  Processing Systems, 17 (2004).

\end{thebibliography}


\begin{thebibliography}{10}

\bibitem{antoulas2017tutorial}
{\sc A.~C. Antoulas, S.~Lefteriu, A.~C. Ionita, P.~Benner, and A.~Cohen}, {\em
  A tutorial introduction to the loewner framework for model reduction}, in
  Model Reduction and Approximation: Theory and Algorithms, vol.~15, SIAM,
  2017, pp.~335--376.

\bibitem{baratchart1991identification}
{\sc L.~Baratchart, M.~Cardelli, and M.~Olivi}, {\em Identification and
  rational {$L^2$} approximation: a gradient algorithm}, Automatica, 27 (1991),
  pp.~413--417.

\bibitem{basak:2022:numerical}
{\sc S.~Basak, S.~Petit, J.~Bect, and E.~Vazquez}, {\em Numerical issues in
  maximum likelihood parameter estimation for gaussian process interpolation},
  in Machine Learning, Optimization, and Data Science. LOD 2021, G.~Nicosia,
  V.~Ojha, E.~La~Malfa, G.~La~Malfa, G.~Jansen, P.~M. Pardalos, G.~Giuffrida,
  and R.~Umeton, eds., Springer International Publishing, Cham, 2022,
  pp.~116--131.

\bibitem{bect:2023:stk}
{\sc J.~Bect, E.~Vazquez, et~al.}, {\em {STK}: a {S}mall ({M}atlab/{O}ctave)
  {T}oolbox for {K}riging. {R}elease 2.8}, 2023,
  \url{https://github.com/stk-kriging/stk/}.

\bibitem{boloix2018complex}
{\sc R.~Boloix-Tortosa, J.~J. Murillo-Fuentes, F.~J. Pay{\'a}n-Somet, and
  F.~P{\'e}rez-Cruz}, {\em Complex {Gaussian} processes for regression}, IEEE
  Trans. Neural Netw. Learn. Syst., 29 (2018), pp.~5499--5511.

\bibitem{boloix2017widely}
{\sc R.~Boloix-Tortosa, J.~J. Murillo-Fuentes, I.~Santos, and
  F.~P{\'e}rez-Cruz}, {\em Widely linear complex-valued kernel methods for
  regression}, IEEE Trans. Signal. Process., 65 (2017), pp.~5240--5248.

\bibitem{Bonizzoni_2020aa}
{\sc F.~Bonizzoni, F.~Nobile, I.~Perugia, and D.~Pradovera}, {\em Least-squares
  {P}ad{\'e} approximation of parametric and stochastic {H}elmholtz maps}, Adv.
  Comput. Math., 46 (2020), pp.~1--28.

\bibitem{Bonyo_2020aa}
{\sc J.~Bonyo}, {\em Reproducing kernels for {Hardy} and {Bergman} spaces of
  the upper half plane}, Commun. Adv. Math. Sci., 3 (2020), pp.~13--23.

\bibitem{burbea:1984:banach}
{\sc J.~Burbea and P.~R. Masani}, {\em Banach and Hilbert Spaces of
  Vector-Valued Functions: Their General Theory and Applications to
  Holomorphy}, Pitman Publishing, Boston, 1984.

\bibitem{CST_2019aa}
{\sc {Dassault Syst{\`e}mes}}, {\em {CST} {STUDIO} {SUITE} 2019},
  \url{https://www.3ds.com/products/simulia/cst-studio-suite}.

\bibitem{deschrijver2008macromodeling}
{\sc D.~Deschrijver, M.~Mrozowski, T.~Dhaene, and D.~De~Zutter}, {\em
  Macromodeling of multiport systems using a fast implementation of the vector
  fitting method}, IEEE Microw. Wireless Compon. Lett., 18 (2008),
  pp.~383--385.

\bibitem{driscoll2014chebfun}
{\sc T.~A. Driscoll, N.~Hale, and L.~N. Trefethen}, {\em Chebfun Guide},
  Pafnuty Publications, Oxford, 2014.

\bibitem{fuhg2021state}
{\sc J.~N. Fuhg, A.~Fau, and U.~Nackenhorst}, {\em State-of-the-art and
  comparative review of adaptive sampling methods for {K}riging}, Arch. Comput.
  Methods Eng.,  (2021), pp.~2689--2747.

\bibitem{Fuhrlander_2020ab}
{\sc M.~Fuhrl{\"a}nder and S.~Sch{\"o}ps}, {\em A blackbox yield estimation
  workflow with {Gaussian} process regression for industrial problems}, J.
  Math. Ind., 10 (2020).

\bibitem{phdGeorg}
{\sc N.~Georg}, {\em Surrogate modeling and uncertainty quantification for
  radio frequency and optical applications}, phd thesis, TU Darmstadt,
  Darmstadt, Germany, 2021.

\bibitem{georg_2024_12601365}
{\sc N.~Georg and J.~Bect}, {\em {Rational kernel-based interpolation for
  complex-valued frequency response functions (matlab code,
  stk-kriging/contrib-cork, v1.0.1)}}.
\newblock Zenodo, July 2024, \url{https://doi.org/10.5281/zenodo.12641323}.

\bibitem{gustavsen2006improving}
{\sc B.~Gustavsen}, {\em Improving the pole relocating properties of vector
  fitting}, IEEE Trans. Power Deliv., 21 (2006), pp.~1587--1592.

\bibitem{gustavsen1999rational}
{\sc B.~Gustavsen and A.~Semlyen}, {\em Rational approximation of frequency
  domain responses by vector fitting}, IEEE Trans. Power Deliv., 14 (1999),
  pp.~1052--1061.

\bibitem{hallemans2022frf}
{\sc N.~Hallemans, R.~Pintelon, B.~Joukovsky, D.~Peumans, and J.~Lataire}, {\em
  {FRF} estimation using multiple kernel-based regularisation}, Automatica, 136
  (2022), p.~110056.

\bibitem{hoffman:1962}
{\sc K.~Hoffman}, {\em Banach spaces of analytic functions}, Prentice Hall,
  Englewood Cliffs, NJ, 1962.

\bibitem{hornikx2015platform}
{\sc M.~Hornikx, M.~Kaltenbacher, and S.~Marburg}, {\em A platform for
  benchmark cases in computational acoustics}, Acta Acust. united Ac., 101
  (2015), pp.~811--820.

\bibitem{jin2002sequential}
{\sc R.~Jin, W.~Chen, and A.~Sudjianto}, {\em On sequential sampling for global
  metamodeling in engineering design}, in International Design Engineering
  Technical Conferences and Computers and Information in Engineering
  Conference, vol.~36223, 2002, pp.~539--548.

\bibitem{kim2009construction}
{\sc B.~Kim, Y.~Lee, and D.-H. Choi}, {\em Construction of the radial basis
  function based on a sequential sampling approach using cross-validation}, J.
  Mech. Sci. Technol., 23 (2009), pp.~3357--3365.

\bibitem{krantz:2001:function}
{\sc S.~G. Krantz}, {\em Function Theory of Several Complex Variables. Second
  Edition}, AMS Chelsea Publishing, Providence, RI, 2001.

\bibitem{lataire2016transfer}
{\sc J.~Lataire and T.~Chen}, {\em Transfer function and transient estimation
  by {Gaussian} process regression in the frequency domain}, Automatica, 72
  (2016), pp.~217--229.

\bibitem{liu2020surrogate}
{\sc Z.~Liu, D.~Lesselier, B.~Sudret, and J.~Wiart}, {\em Surrogate modeling
  based on resampled polynomial chaos expansions}, Reliab. Eng. Syst. Saf., 202
  (2020), p.~107008.

\bibitem{meinshausen2010stability}
{\sc N.~Meinshausen and P.~B{\"u}hlmann}, {\em Stability selection}, J. R.
  Stat. Soc., B: Stat. Methodol., 72 (2010), pp.~417--473.

\bibitem{micchelli:2005:vector}
{\sc C.~A. Micchelli and M.~Pontil}, {\em On learning vector-valued functions},
  Neural Comput., 17 (2005), pp.~177--204.

\bibitem{miller1969complex}
{\sc K.~S. Miller}, {\em Complex {Gaussian} processes}, SIAM Rev., 11 (1969),
  pp.~544--567.

\bibitem{nakatsukasa2018aaa}
{\sc Y.~Nakatsukasa, O.~S{\`e}te, and L.~N. Trefethen}, {\em The {AAA}
  algorithm for rational approximation}, SIAM J. Sci. Comput., 40 (2018),
  pp.~A1494--A1522.

\bibitem{nobile2020non}
{\sc F.~Nobile and D.~Pradovera}, {\em Non-intrusive double-greedy parametric
  model reduction by interpolation of frequency-domain rational surrogates},
  arXiv:2008.10864,  (2020).

\bibitem{Paulsen_2016aa}
{\sc V.~I. Paulsen and M.~Raghupathi}, {\em An introduction to the theory of
  reproducing kernel {Hilbert} spaces}, vol.~152, Cambridge University Press,
  2016.

\bibitem{petit:2023:parameter}
{\sc S.~Petit, J.~Bect, P.~Feliot, and E.~Vazquez}, {\em Parameter selection in
  {G}aussian process interpolation: an empirical study of selection criteria},
  SIAM/ASA Journal on Uncertainty Quantification, 11 (2023).

\bibitem{picinbono:1996}
{\sc B.~Picinbono}, {\em Second-order complex random vectors and normal
  distributions}, IEEE Trans. Signal Process, 44 (1996), pp.~2637--2640.

\bibitem{picinbono1995widely}
{\sc B.~Picinbono and P.~Chevalier}, {\em Widely linear estimation with complex
  data}, IEEE Trans. Signal Process., 43 (1995), pp.~2030--2033.

\bibitem{pillonetto2010new}
{\sc G.~Pillonetto and G.~De~Nicolao}, {\em A new kernel-based approach for
  linear system identification}, Automatica, 46 (2010), pp.~81--93.

\bibitem{pradovera2023toward}
{\sc D.~Pradovera}, {\em Toward a certified greedy {L}oewner framework with
  minimal sampling}, Advances in Computational Mathematics, 49 (2023).

\bibitem{rodriguez2020p}
{\sc A.~C. Rodriguez and S.~Gugercin}, {\em The {p-AAA} algorithm for data
  driven modeling of parametric dynamical systems}, SIAM Journal on Scientific
  Computing, 45 (2023).

\bibitem{Roemer_2021aa}
{\sc U.~R{\"o}mer, M.~Bollh{\"o}fer, H.~Sreekumar, C.~Blech, and S.~Langer},
  {\em An adaptive sparse grid rational arnoldi method for uncertainty
  quantification of dynamical systems in the frequency domain}, Int. J. Numer.
  Meth.,  (2021), pp.~5487--5511.

\bibitem{schneider2023sparse}
{\sc F.~Schneider, I.~Papaioannou, and G.~M{\"u}ller}, {\em Sparse bayesian
  learning for complex-valued rational approximations}, International Journal
  for Numerical Methods in Engineering, 124 (2023), pp.~1721--1747.

\bibitem{tisseur2001quadratic}
{\sc F.~Tisseur and K.~Meerbergen}, {\em The quadratic eigenvalue problem},
  SIAM Rev., 43 (2001), pp.~235--286.

\bibitem{treviso2021multiple}
{\sc F.~Treviso, R.~Trinchero, and F.~G. Canavero}, {\em Multiple delay
  identification in long interconnects via {LS-SVM} regression}, IEEE Access, 9
  (2021), pp.~39028--39042.

\bibitem{ziegelwanger2017pac}
{\sc H.~Ziegelwanger and P.~Reiter}, {\em The {PAC-MAN} model: Benchmark case
  for linear acoustics in computational physics}, J. Comput. Phys., 346 (2017),
  pp.~152--171.

\end{thebibliography}


\cleardoublepage
\renewcommand\appendixname{}
\phantomsection
\addcontentsline{toc}{section}{SUPPLEMENTARY MATERIAL}%
\setcounter{section}{0} \renewcommand\thesection {SM\arabic{section}}
\setcounter{figure}{0}  \renewcommand\thefigure  {SM\arabic{figure}}
\makeatletter 
\renewcommand\section{\@startsection{section}{1}{.25in}%
                                   {1.3ex \@plus .5ex \@minus .2ex}%
                                   {-.5em \@plus -.1em}%
                                   {\reset@font\normalsize\bfseries}}
\renewcommand\subsection{\@startsection{subsection}{2}{.25in}%
                                     {1.3ex\@plus .5ex \@minus .2ex}%
                                     {-.5em \@plus -.1em}%
                                     {\reset@font\normalsize\bfseries}}
\renewcommand\subsubsection{\@startsection{subsubsection}{3}{.25in}%
                                     {1.3ex\@plus .5ex \@minus .2ex}%
                                     {-.5em \@plus -.1em}%
                                     {\reset@font\normalsize\bfseries}}
\makeatother
\noindent\rule{\textwidth}{3pt}\\%
\begin{center}
  \textbf{SUPPLEMENTARY MATERIAL}
\end{center}
\vspace{5pt}%
\noindent\rule{\textwidth}{3pt}%

\vspace{5mm}

\section{Introduction}
This supplementary material is structured in the following way. %
We first present a non-intrusive implementation of the new method presented in
the paper in Section~\ref{sec:non-intrusive}. 
Section~\ref{sec:additional_results} reports additional numerical results and details on the examples used in the main text. In particular, some investigations related to the estimation of the kernel parameter $\alpha$ are given in Section~\ref{sec:alpha}, a partial fraction representation for the circuit model is derived in Section~\ref{sec:circuit}, Section~\ref{sec:kmax_study} discusses the choice and influence of the maximum number of poles in the mean model $K_{\mathrm{max}}$ and results for a spiral antenna, which are comparable to the PAC-MAN model shown in the main paper, are contained in Section~\ref{sec:add-num-results}. %
Finally, some theoretical results regarding circular complex/real RKHS
are collected in Section~\ref{sec:circular}.

\section{Non-intrusive implementation}
\label{sec:non-intrusive}

\subsection{Zero-mean case}

The main idea is to construct an isomorphic real-valued GP
$\tilde g(\tilde{\mathbf x})\sim \operatorname{GP}(0,\tilde k)$ on an
\textit{augmented input space}
$\tilde{\mathbf x}\in \bigl(\mathbb R^n\times \{0,1\}\bigr)$, s.t.,
\begin{equation}
  \begin{aligned}
    \tilde g(\begin{bmatrix}\mathbf x&0\end{bmatrix}) &= \operatorname{Re}[g(\mathbf x)],\\
    \tilde g(\begin{bmatrix}\mathbf x&1\end{bmatrix}) &= \operatorname{Im}[g(\mathbf x)].
  \end{aligned}\label{eq:idea}
\end{equation}
The augmented training data
$\bigl(\tilde {\mathbf x},\tilde y\bigr) \in \bigl(\mathbb R^n \times
\{0,1\}\bigr) \times \mathbb R$ is for each observation
$\bigl(\mathbf x^{(i)}, y^{(i)}\bigr)\in \mathbb R^n\times\mathbb C$
obtained as:
\begin{align*}
  \tilde{\mathbf x}^{(i,1)} &= \begin{bmatrix}
    \mathbf x^{(i)} & 0
  \end{bmatrix},~~\tilde {y}^{(i,1)} = 
                      \operatorname{Re}[y^{(i)}],\\
  \tilde{\mathbf x}^{(i,2)} &= \begin{bmatrix}
    \mathbf x^{(i)} & 1
  \end{bmatrix},~~\tilde {y}^{(i,2)} = 
                      \operatorname{Im}[y^{(i)}].
\end{align*}
The \textit{new} covariance function $\tilde k$ can be derived by
enforcing~\eqref{eq:idea},
\begin{align*}
  \tilde k(\tilde {\mathbf x}, \tilde {\mathbf x}')&=\tilde k([\mathbf x~j],[\mathbf x'~j'])
  =\begin{cases}
    \frac 12\operatorname{Re}[k(\mathbf x,\mathbf x')+c(\mathbf x,\mathbf x')] &j=j'=0\\
    \frac 12\operatorname{Re}[k(\mathbf x,\mathbf x')-c(\mathbf x,\mathbf x')] &j=j'=1\\
    \frac 12\operatorname{Im}[-k(\mathbf x,\mathbf x')+c(\mathbf x,\mathbf x')] &j=0,j'=1\\
    \frac 12\operatorname{Im}[k(\mathbf x,\mathbf x')+c(\mathbf x,\mathbf x')] &j=1,j'=0
  \end{cases},
\end{align*}
Note that this approach requires to define the modified covariance
function $\tilde k$, however, no (other) internal functions of
existing GP implementations need to be able to cope with complex
numbers, which is why we refer to the implementation as
\emph{non-intrusive}.

\subsection{Linear model in the mean function}

Consider now the superposition
\begin{equation*}
  \underline g(\mathbf x) = g(\mathbf x) + \mathbf h(\mathbf x)^\mathrm{T} \mathbf b
\end{equation*}
of a mean-free (real/)complex Gaussian Process
$g(\mathbf x)\sim \operatorname{CGP}(0, k, c)$ and a complex linear
model, where $h(\mathbf x):\mathbb R^n\rightarrow \mathbb C^m$ denote
explicit basis functions and $\mathbf b\in\mathbb C^m$ the
corresponding coefficients.
Define the augmented process %
\begin{equation*}
  \tilde{\underline g}(\tilde{\mathbf x}) =
  \tilde g(\tilde{\mathbf x})+\tilde{\mathbf h}(\tilde{\mathbf x})^\mathrm{T}\tilde{\mathbf b},  
\end{equation*}
where $\tilde g(\tilde{\mathbf x})~\sim \mathrm{GP}(0, \tilde k),$
$\tilde{\mathbf b}\in \mathbb R^{2m}$,
$\tilde{\mathbf h}(\tilde{\mathbf x}):\mathbb
R^n\times\{0,1\}\rightarrow \mathbb R^{2m}$ and we require, similarly
as in the last subsection, that
\begin{align*}
  \tilde{\underline g}([\mathbf x~0])&=\operatorname{Re}[\underline g(\mathbf x)],\\
  \tilde{\underline g}([\mathbf x~1])&=\operatorname{Im}[\underline g(\mathbf x)].
\end{align*}
Incorporating~\eqref{eq:idea}, we can conclude that
\begin{align*}
  \tilde{\mathbf h}([\mathbf x~0])^\mathrm{T}\tilde{\mathbf b}&=\operatorname{Re}[\mathbf h(\mathbf x)^\mathrm{T} \mathbf b]=\operatorname{Re}[\mathbf h(\mathbf x)]^\mathrm{T}\operatorname{Re}[\mathbf b]-\operatorname{Im}[\mathbf h(\mathbf x)]^\mathrm{T}\operatorname{Im}[\mathbf b]\\
  \tilde{\mathbf h}([\mathbf x~1])^\mathrm{T}\tilde{\mathbf b}&=\operatorname{Im}[\mathbf h(\mathbf x)^\mathrm{T} \mathbf b]=\operatorname{Im}[\mathbf h(\mathbf x)]^\mathrm{T}\operatorname{Re}[\mathbf b]+\operatorname{Re}[\mathbf h(\mathbf x)]^\mathrm{T}\operatorname{Im}[\mathbf b].
\end{align*}
needs to be fulfilled.
This can be achieved by setting 
\begin{equation*}
  \tilde{\mathbf h}(\tilde{\mathbf x})=\tilde{\mathbf h}([\mathbf x~j])=\begin{bmatrix}
    \begin{cases}
      +\operatorname{Re}[\mathbf h(\mathbf x)] & j=0\\
      +\operatorname{Im}[\mathbf h(\mathbf x)] & j=1
    \end{cases}&\hspace{2em}
    \begin{cases}
      -\operatorname{Im}[\mathbf h(\mathbf x)] & j=0\\
      +\operatorname{Re}[\mathbf h(\mathbf x)] & j=1
    \end{cases}
  \end{bmatrix}^{\mathrm{T}},
\end{equation*}
which leads to the coefficients vector
\begin{equation*}
  \tilde{\mathbf b} = \begin{bmatrix}
    \operatorname{Re}[\mathbf b]\\
    \operatorname{Im}[\mathbf b]
  \end{bmatrix}.
\end{equation*}

\section{Complements to the numerical results}
\label{sec:additional_results}

\subsection{Estimation of alpha for low order rational functions}
\label{sec:alpha}
We conduct an additional study to investigate whether the selected values of $\alpha$ provide insights into the properties of the approximated function. We employ the Szegö kernel-based approximation using log-likelihood maximization as described in the main part of the paper. We consider low order rational functions, as introduced in Section 2 of the paper, but vary the real part of the poles. The associated functions 
\begin{equation}\label{eq:Frat_alpha}
	F_{\mathrm{rat}, \beta}(\i\omega; \beta)
	= \frac 1 {\i\omega-(-\beta)}
	+ \frac {0.5}{\i\omega-(-\beta-0.5\i)}
	+ \frac {0.5}{\i\omega-(-\beta+0.5\i)},\;
	\omega\in[0,1],
\end{equation}
where $F_{\mathrm{rat}, \beta}(\i\omega; \beta) \in H^2_\mathrm{sym}(\Gamma_{\beta+\epsilon})$,
are plotted in Fig.~\ref{fig:rational_functions}, along with the respective selected values of $\alpha$, where we have always employed $n=20$ training points. It can be seen that the selected values of $\alpha$ are quite close to the real part of the poles of the function for the case with and without the Hermitian symmetry condition. Unfortunately, this relation is not easy to reveal or even investigate for more complicated examples.

\begin{figure}
	\begin{tikzpicture}
	\begin{axis}[grid, 	xlabel={$\omega$}, title={Real Part},
	width=.4\textwidth, height=.3\textwidth, ylabel near ticks,
	legend style = {font=\small}, legend pos = north east, xmin=0, xmax=1]
	\addplot[TUDa-1b, very thick, no marks] table[x=x, y=y_real_alpha0.25,
	col sep = comma]{images/illustration_rational_functions.csv};
	\addplot[TUDa-9b, very thick, no marks] table[x=x, y=y_real_alpha0.2,
	col sep = comma]{images/illustration_rational_functions.csv};
	\addplot[TUDa-4d, very thick, no marks] table[x=x, y=y_real_alpha0.15,
	col sep = comma]{images/illustration_rational_functions.csv};
	\addplot[TUDa-10b, very thick, no marks] table[x=x, y=y_real_alpha0.1,
	col sep = comma]{images/illustration_rational_functions.csv};
	\addplot[TUDa-7b, very thick, no marks] table[x=x, y=y_real_alpha0.05,
	col sep = comma]{images/illustration_rational_functions.csv};
	\end{axis}
	\end{tikzpicture}
	\begin{tikzpicture}
	\begin{axis}[grid, 	xlabel={$\omega$}, title={Imaginary Part},
	width=.4\textwidth, height=.3\textwidth, ylabel near ticks,
	legend style = {font=\small}, legend pos = outer north east, xmin=0, xmax=1]
	\addplot[TUDa-1b, very thick, no marks] table[x=x, y=y_imag_alpha0.25,
	col sep = comma]{images/illustration_rational_functions.csv};
	\addplot[TUDa-9b, very thick, no marks] table[x=x, y=y_imag_alpha0.2,
	col sep = comma]{images/illustration_rational_functions.csv};
	\addplot[TUDa-4d, very thick, no marks] table[x=x, y=y_imag_alpha0.15,
	col sep = comma]{images/illustration_rational_functions.csv};
	\addplot[TUDa-10b, very thick, no marks] table[x=x, y=y_imag_alpha0.1,
	col sep = comma]{images/illustration_rational_functions.csv};
	\addplot[TUDa-7b, very thick, no marks] table[x=x, y=y_imag_alpha0.05,
	col sep = comma]{images/illustration_rational_functions.csv};
	\legend{$F_{\mathrm{rat}, \beta}(\i\omega; 0.25)$, $F_{\mathrm{rat}, \beta}(\i\omega; 0.2)$, $F_{\mathrm{rat}, \beta}(\i\omega; 0.15)$, $F_{\mathrm{rat},\beta}(\i\omega; 0.1)$,  $F_{\mathrm{rat},\beta}(\i\omega; 0.05)$}
	\end{axis}
	\end{tikzpicture}\\
	\centering
	\begin{tikzpicture}
	\begin{axis}[grid, 	xlabel={$\beta$}, ylabel={Estimated $\alpha$},
	width=.45\textwidth, height=.33\textwidth, ylabel near ticks,
	legend style = {font=\small}, legend pos = outer north east, xmin=0.05, xmax=0.25, ymin=0.05, ymax=0.25, xtick distance = 0.05, ytick distance = 0.05, x tick label style={/pgf/number format/.cd,fixed,precision=2}, y tick label style={/pgf/number format/.cd,fixed,precision=2}, mark options={solid}]
	\addplot[TUDa-1b, very thick, dashed, mark=*] table[x=alpha, y=alpha_est_PCov,
	col sep = comma]{images/alpha_estimation.csv};
	\addplot[TUDa-9b, very thick, dotted, mark=square*] table[x=alpha, y=alpha_est_NoPCov,
	col sep = comma]{images/alpha_estimation.csv};
	\legend{{Szegö with pseudo-kernel}, {Szegö with $c=0$}}
	\end{axis}
	\end{tikzpicture}
	\caption{Top: Illustration of low order complex rational functions with varying real part of the poles. Bottom: Selected values of $\alpha$ based on $n=20$ training points, either using the pseudo-kernel enforcing Hermitian symmetry or $c=0$.}
	\label{fig:rational_functions}
\end{figure}

\subsection{Partial fraction representation of RLC circuit}
\label{sec:circuit}
The residues $c_i, c_i^*$ and poles $a_i, a_i^*$ of the partial fraction representation of the electric circuit admittance  
\begin{equation}
  Y(s) = \sum_{i=1}^N \frac {c_i}{s-a_i}+\frac
  {c_i^*}{s -a_i^*},
\end{equation}
are given as
\begin{align}
a_i &= \frac {-R_i}{2L_i}+ \i\sqrt{\frac 1 {L_iC_i}-\Bigl(\frac {R_i}{2L_i}\Bigr)^2},\\
c_i &= \frac {a_i} {L(a_i-a_i^*)} = \frac {\sqrt{\frac 1 {L_iC_i}-\Bigl(\frac {R_i}{2L_i}\Bigr)^2}+\frac {R_i}{2L_i}\i}{2L_i\sqrt{\frac 1 {L_iC_i}-\Bigl(\frac {R_i}{2L_i}\Bigr)^2}},
\end{align}
where we assumed an underdamped system, i.e. \begin{equation}\frac {R_i}2\sqrt{\frac{C_i}{L_i}}<1,\label{[eq:cond_underdamped}\end{equation} which implies that the argument of the square roots is positive.

\subsection{The choice of $K_\mathrm{max}$}
\label{sec:kmax_study}

For the experiments contained in the paper we limit the maximum number of poles pairs to $K_{\max} = \min\{5,\lfloor n/4 \rfloor\}$, where the limit of $5$ is considered as an arbitrary \textit{small} number, based on the assumption that the approximated functions have a small number of dominant poles. Note that increasing $K_\mathrm{max}$ would increase the computational cost of the model selection. In Fig.~\ref{fig:study_wrt_kmax}, we demonstrate that the particular choice of this value is indeed not very important for the particular random circuit realization with admittance $Y_2$ and $n=50$ training points, by computing repeated approximations based on different values of $K_\mathrm{max}$. It can be observed that for $K_\mathrm{max} > 2$ the different selected approximations have a similar error magnitude. The black line additionally indicates the selected number of pole pairs. Note that, even when the same number of pole pairs is chosen for different $K_\mathrm{max}$, the respective approximations are not necessarily exactly the same, as different initial values are employed for the gradient-based optimization as explained in Section 3.2 of the paper.

\begin{figure}
	\centering
	\begin{tikzpicture}
	\begin{axis}[grid, 	xlabel={Maximum number of pole pairs $K_\mathrm{max}$.}, width=.8\textwidth, height=.45\textwidth, ylabel near ticks,
	legend style = {font=\small}, legend pos = outer north east, xmin=0, xmax=15, ymin=0, ymax=7.3, xtick distance=1, ytick distance=1, mark options={solid}]
	\addplot[TUDa-1b, very thick, dashed, mark=*] table[x=max_num_pole_pairs, y=rmse,
	col sep = comma]{images/error_circuit_approx_different_max_pole_pairs.csv};
	\addplot[TUDa-9b, very thick, dashed, mark=diamond*] table[x=max_num_pole_pairs, y=max_err,
	col sep = comma]{images/error_circuit_approx_different_max_pole_pairs.csv};
	\addplot[black, thick, dotted, mark=triangle*] table[x=max_num_pole_pairs, y=chosen_num_pole_pairs,
	col sep = comma]{images/error_circuit_approx_different_max_pole_pairs.csv};
	\legend{RMSE, Max. Error, Selected $k$}
	\end{axis}
	\end{tikzpicture}
	\caption{Approximation accuracy for the random circuit realization shown in Fig. 7 of the paper with $n=50$ training points and varying maximum number of pole pairs $K_\mathrm{max}$. Additionally, the respective number of pole pairs as chosen by the model selection is indicated by black triangles.}
	\label{fig:study_wrt_kmax}
\end{figure}

\subsection{Additional numerical example}
\label{sec:add-num-results}

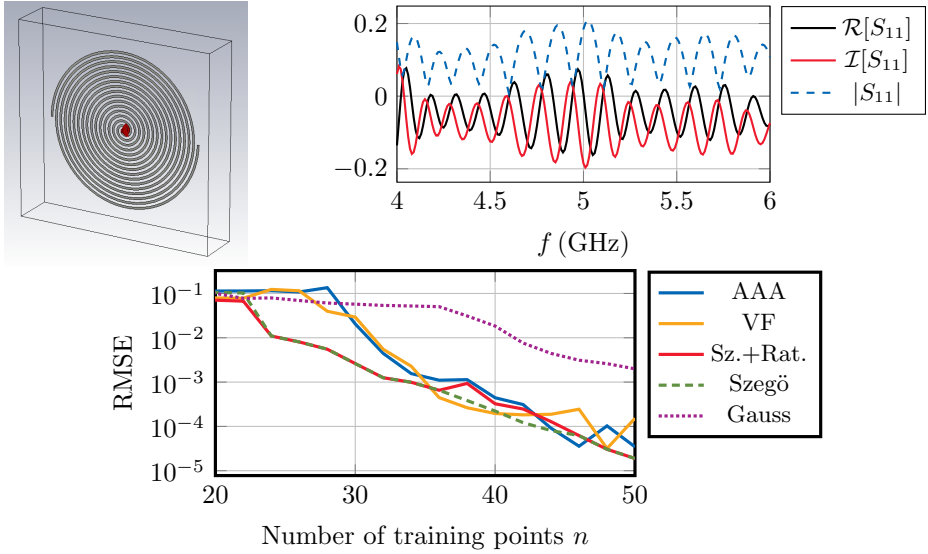
\begin{figure}
	\begin{subfigure}[b]{.31\textwidth}
		\centering
		\includegraphics[width=.8\textwidth]{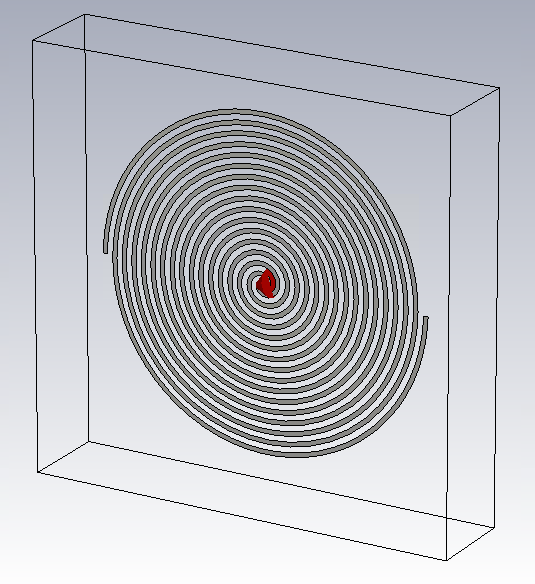}
	\end{subfigure}\hfill
	\begin{subfigure}[b]{.67\textwidth}
		\centering
		\begin{tikzpicture}
		\begin{axis}[grid,xlabel={$f$\,(GHz)},
		width=.75\linewidth, height=.45\linewidth,  ylabel near ticks, legend style = {font=\small}, legend pos = outer north east, xmin = 4, xmax=6, every axis plot/.append style={very thick}]
		\addplot[black,thick, no marks] table[x=x, y=y_ref_real, col sep = comma]{images/illustration_Spiral.csv};
		\addplot[TUDa-9b, thick, no marks] table[x=x, y=y_ref_imag, col sep = comma]{images/illustration_Spiral.csv};
		\addplot[TUDa-1b, dashed, thick, no marks] table[x=x, y=y_ref_abs, col sep = comma]{images/illustration_Spiral.csv};
		\legend{$\mathcal R[S_{11}]$, $\mathcal I[S_{11}]$, $|S_{11}|$}
		\end{axis}
		\end{tikzpicture}
	\end{subfigure}
	\begin{subfigure}[b]{1\textwidth}
		\centering
		\begin{tikzpicture}
		\begin{semilogyaxis}[grid, xlabel={Number of training points $n$}, ylabel={RMSE}, width=.55\textwidth, height=.331\textwidth, legend pos = outer north east,  legend style = {font=\small}, ytick = {1e-1, 1e-2, 1e-3, 1e-4, 1e-5}, very thick, no marks, xmin = 20, xmax=50, every axis plot/.append style={very thick}]
		\addplot[TUDa-1b] table[x=N, y=RMSE, col sep = comma]{images/Spiral_AAA.csv};
		\addplot[TUDa-7b] table[x=N, y=RMSE, col sep = comma]{images/Spiral_VF.csv};
		\addplot[TUDa-9b] table[x=N, y=RMSE, col sep = comma]{images/Spiral_Adap.csv};
		\addplot[TUDa-4d, densely dashed] table[x=N, y=RMSE, col sep = comma]{images/Spiral_Szego.csv};
		\addplot[TUDa-10b, densely dotted] table[x=N, y=RMSE, col sep = comma]{images/Spiral_GaussSep25.csv};
		\legend{AAA, VF, Sz.+Rat.,Szegö, Gauss}
		\end{semilogyaxis}
		\end{tikzpicture}
	\end{subfigure}
	\caption{Top left: Spiral antenna model, taken from CST
          Microwave Studio \cite{CST_2019aa}. Top right: Complex
          frequency response function $S_{11}$. Bottom: Convergence
          study w.r.t.\ the number of training points.}
	\label{fig:spiral}
\end{figure}
The model is a spiral antenna, depicted in Figure~\ref{fig:spiral},  where we consider the reflection coefficient $S_{11}$ on a frequency range of [\SI{4}{GHz},\,\SI{6}{GHz}] as quantity of interest. The data sets are obtained using the boundary element method in CST Microwave Studio \cite{CST_2019aa}. The results are qualitatively the same as for the PAC-MAN model, see  Fig.~\ref{fig:spiral} (bottom).

\section{Complements for the theoretical section}

\subsection{Circular complex/real RKHSs}
\label{sec:circular}    

\begin{definition}
  We say that a complex/real RKHS is \emph{circular} if it has a
  vanishing pseudo-kernel.
\end{definition}

The terms ``proper'' or ``strictly complex'' are also sometimes used
instead of ``circular'', in the statistics and signal processing
literature, for the case where the
pseudo-covariance of a complex Gaussian random vector or function
vanishes (see, e.g., \cite{picinbono:1996, boloix2018complex}).

\begin{theorem} \label{thm:circular}
  Let $H_\Cset$ denote a complex RKHS with kernel~$k_0$, and let $H$
  denote the complex/real RKHS obtained by considering~$H_\Cset$ as a
  real vector space, endowed with the inner product:
  $\left<f, g\right> \mapsto \Re\left( \left<f, g\right>_\Cset \right)$. %
  Then $H$ is the circular complex/real RKHS with complex kernel
  $k = 2 k_0$.  
\end{theorem}

Since a complex/real RKHS is uniquely characterized by its $(k, c)$
pair, the converse holds as well: %
given a circular complex/real RKHS~$H$ with complex kernel~$k$, there
is a unique complex RKHS~$H_\Cset$ (namely, the complex RKHS with
kernel $k_0 = \frac{1}{2} k$) such that $H$ is obtained from~$H_\Cset$
as in Theorem~\ref{thm:circular}.

\begin{proof}
  The main idea is already included in the proof of Theorem~2.14 of
  the article, but we give here a slightly more detailed version. %
  Let $\varphi_\re$ and~$\varphi_\im$ denote the real and imaginary
  evaluation kernels of~$H$. %
  Then, for all~$f \in H$ and~$s \in \Sset$,
  \begin{equation*}
    \left< f,\, k_0(\cdot,s) \right>
    \;=\; \Re\left( \left< f,\, k_0(\cdot,s) \right> \right)
    \;=\; \Re\left( f(s) \right)
  \end{equation*}
  and
  \begin{align*}
    \left< f,\, \i k_0(\cdot,s) \right>
    & \;=\; \Re\left( \left< f,\, \i k_0(\cdot,s) \right> \right)
      \;=\; \Re\left( -\i\, \left< f,\, k_0(\cdot,s) \right> \right)\\
    & \;=\; \Re\left( -i\ f(s) \right)
      \;=\; \Im\left( f(s) \right),
  \end{align*}
  which proves that~$\varphi_\re = k_0$ and~$\varphi_\im = \i k_0$. %
  The complex kernel and pseudo-kernel of~$H$ are thus given by
  \begin{align*}
    k & \;=\; \varphi_\re - \i \varphi_\im \;=\; 2 k_0,\\
    c & \;=\; \varphi_\re + \i \varphi_\im \;=\; 0.
  \end{align*}
\end{proof}

\subsection{A relation between the Szegö and rational quadratic kernels}
\label{sec:relation-szego-rq}

Consider the Szegö kernel on $H^2(\Gamma_\alpha)$:
\begin{align*}
  k_\alpha\left( s, s_0 \right)
  & \;=\; \frac 1 {2\pi\left( 2\alpha + s + s_0^* \right)}\\
  & \;=\; \frac{1}{2\pi}\, \frac{%
    \left( 2\alpha + x + x_0 \right) - \i \left( y - y_0 \right)}{%
    \left( 2\alpha + x + x_0 \right)^2 + \left( y - y_0 \right)^2},
\end{align*}
where $s=x+\i y$, $s_0 = x_0 + \i y_0 \in \Gamma_\alpha$.
In the circular case, the corresponding kernels for the real and
imaginary parts are given by:
\begin{equation*}
  k_\re (s, s_0) \;=\; k_\im (s, s_0)
  \;=\; \frac{1}{4\pi}\, \frac{2\alpha + x + x_0}{%
    \left( 2\alpha + x + x_0 \right)^2 + \left( y - y_0 \right)^2},
\end{equation*}
For a fixed value of~$x = x_0 > -\alpha$, this is of the form
\begin{equation*}
  (y, y_0) \;\mapsto\; \frac{1}{4\pi}\, \frac{A}{%
    A^2 + \left( y - y_0 \right)^2},
  \qquad \text{with } A = 2\alpha + x + x_0 > 0,
\end{equation*}
which is a special case of the so-called \emph{rational quadratic}
kernel (see, e.g.,~\citeSM{rasmussen:2006:gpml, sollich:2004:using}),
also called \emph{generalized inverse multiquadric} kernel (see,
e.g.,~\citeSM{hu:1998:collocation, bozzini:2013:generalized}).

\phantomsection
\bibliographystyleSM{siamplain}
\bibliographySM{references}

\end{document}